\documentclass[journal]{IEEEtran}
\usepackage{amsmath, amssymb, cite, tikz, graphicx, xspace, mathrsfs, mathtools, psfrag, chemarrow, subfigure,color,soul,graphicx,diagbox}
\usepackage{enumitem}
\usepackage{bbm}
\usepackage[utf8]{inputenc}
\usepackage{booktabs}
\usepackage{amscd}
\usepackage{amsmath}
\usepackage{amssymb}
\usepackage{amsthm}

\usepackage{epsfig}
\usepackage{verbatim}
\usepackage{graphicx}
\usepackage{amsthm}
\usepackage{color}
\usepackage{ multirow }
\usepackage{algorithm}
\usepackage[noend]{algpseudocode}
\usepackage{pgfplots}
\usetikzlibrary{patterns}
\usetikzlibrary{angles} 
\usepackage{tikz}
\usetikzlibrary{shapes, arrows, decorations.markings, arrows.meta}
\usetikzlibrary{patterns} 
\usetikzlibrary{arrows,automata,positioning,chains}
\usetikzlibrary{quotes,angles}
\newtheorem{remark}{Remark}
\makeatletter
\newcommand{\vast}{\bBigg@{3.25}}
\newcommand{\Vast}{\bBigg@{4}}
\makeatother
\newtheorem{theorem}{Theorem}
\newtheorem{lemma}[theorem]{Lemma}
\begin{document}
	
	\title{State-aware Real-time Tracking and Remote Reconstruction of a Markov Source}
	
	\author{Mehrdad Salimnejad, Marios Kountouris, Nikolaos Pappas
		\thanks{M. Salimnejad and N. Pappas are with the Department of Computer and Information Science Linköping University, Sweden, email: \{\texttt{mehrdad.salimnejad, nikolaos.pappas\}@liu.se}. M. Kountouris is with the Communication Systems Department at EURECOM, Sophia-Antipolis, France, email: \texttt{marios.kountouris@eurecom.fr}. \\ The work of M. Salimnejad and N. Pappas has been supported in part by the Swedish Research Council (VR), ELLIIT, Zenith, and the European Union (ETHER, 101096526). The work of M. Kountouris has received funding from the European Research Council (ERC) under the European Union’s Horizon 2020 Research and Innovation programme (Grant agreement No. 101003431).}}
	
	\maketitle
	\begin{abstract}
		\par The problem of real-time remote tracking and reconstruction of a two-state Markov process is considered here. A transmitter sends samples from an observed information source to a remote monitor over an unreliable wireless channel. The receiver, in turn, performs an action according to the state of the reconstructed source. We propose a state-aware randomized stationary sampling and transmission policy which accounts for the importance of different states of the information source, and their impact on the goal of the communication process. 
		We then analyze the performance of the proposed policy, and compare it with existing goal-oriented joint sampling and transmission policies, with respect to a set of performance metrics. Specifically, we study the real-time reconstruction error, the cost of actuation error, the consecutive error, and a new metric, coined importance-aware consecutive error. In addition, we formulate and solve a constrained optimization problem that aims to obtain the optimal sampling probabilities that minimize the average cost of actuation error. Our results show that in the scenario of constrained sampling generation, the optimal state-aware randomized stationary policy outperforms all other sampling policies for fast evolving sources, and, under certain conditions, for slowly varying sources. Otherwise, a semantics-aware policy performs better only when the source is slowly varying. 
	\end{abstract}	
	
	\section{Introduction}
	Today's communication networks are in a transitional phase to supporting cyber-physical and interactive critical systems, which are key enablers for a plethora of new services and applications, such as autonomous transportation, industrial robotics, telehealth, and environmental monitoring. Emerging real-time autonomous systems, empowered with networked agents with advanced processing and learning capabilities, are expected to take advantage of network and sensing data and transform both human and digital decision making. Nevertheless, the realization of this euphoric vision hinges upon networks’ ability to timely and effectively gather, analyze, and transport vast new sources of data. 
	As a step in that direction, a radically new approach, which accounts for the \emph{semantics of information}, defined as the importance and the goal-oriented utility of data exchanged in a network, has emerged. Reconsidering the entire communication process under the prism of semantics of information is instrumental in transforming the way we generate, transmit, and reconstruct data in time-sensitive and data-intensive communication systems. Anthony Ephremides is among the very first who proposed and advocated for the concept of semantics of information, laying the foundation stones of goal-oriented semantic communications. A highly relevant yet challenging problem in this context is to design joint source sampling, transmission, and reconstruction techniques, which consider the dynamics of the information source and enable real-time remote tracking with the objective of actuation. 
	
	Most prior work on remote tracking to date has mainly focused on proposing sampling or scheduling policies aiming to minimize estimation error or mean square error, letting the significance and the usefulness of the generated and transmitted information with respect to the application-driven goal and context aside. In contrast to these works, in this paper, we propose a new state-aware sampling and transmission policy and introduced a new importance-aware error metric, unearthing the prominent role of having different action probabilities for different states. 
	
	\subsection{Related work}
	\par The problem of scheduling in event-triggered estimation has been considered in \cite{shi2011sensor,wu2012event,meng2012optimal,wu2013can,trimpe2014event,leong2016sensor}, where a sensor observes the state of a process and transmits it to the receiver only when certain events occur. Optimal sampling and transmission policies for noiseless communication channels are proposed in \cite{imer2005optimal,nayyar2013optimal,chakravorty2014optimal}. The study in \cite{imer2005optimal} considers sequential estimation with limited information, where an observer sequentially observes a stochastic process and sends the resulting sample to a receiver over a noiseless communication channel. The authors in \cite{nayyar2013optimal} study a remote estimation problem in a noiseless communication system in the presence of an energy harvesting sensor and a remote estimator. \cite{chakravorty2014optimal} presents an optimal threshold transmission policy for a noiseless communication system where a sensor observes a first-order Markov process and transmits the sample to the receiver. 
	
	The work \cite{shi2012scheduling} proposes an optimal transmission strategy in two sensor-assisted Gauss-Markov systems, extended to multiple sensors and processes in \cite{wu2018optimal}. The work \cite{chakravorty2019remote} analyzes the optimal estimation and transmission policies for remote estimation over time-varying packet drop channels. This study involves a scenario representing the information source as a finite-state Markov chains and first-order auto-regressive processes. The authors in \cite{chakravorty2015distortion} and \cite{chakravorty2016fundamental}  study the fundamental limits and trade-offs of remote estimation of Markov processes under communication constraints. 
	
	Optimal sampling and remote estimation for monitoring real-time stochastic processes is studied in \cite{sun2019sampling,ornee2021sampling,GuoKostina2022,hui2022real}. The problem of estimating the current state of a dynamic process using previous measurements and a linear time-invariant discrete-time (LTI) model of the process is investigated in \cite{lipsa2011remote,huang2019retransmit,pezzutto2022transmission}. The main objective of the aforementioned studies is to present sampling and transmission strategies that minimize estimation errors, disregarding the importance of information with respect to its utilization. Metrics that capture the semantics and effectiveness of information, leveraging synergies between data processing, information transmission, and signal reconstruction have recently been introduced in \cite{kountouris2021semantics,pappas2021goal,tolga21SP,lan2021semantic,Qin22arxiv,PetarProc2022,stavrou2022rate,stamatakis2022semantics,jayanth23, GunduzJSAC23,cocco2023remote,salimnejad2023ICC,salimnejad2023TCOM,fountoulakis2023goal}.
	
	\subsection{Contributions}
	In this work, we consider the problem of real-time remote tracking of an information source in a time slotted communication system. A sampler performs sampling of a two-state Markov process, and then the transmitter sends the sample in the form of packets to a remote receiver over an unreliable wireless channel. Then, the real-time reconstruction of the information source is performed at the receiver based on the successfully received samples. The system is considered to be in a synced state if the source state matches the state of the reconstructed source, otherwise the system is in an erroneous state. Furthermore, the receiver performs a specific action according to the estimated state of the information source. This paper extends the results of \cite{pappas2021goal,salimnejad2023TCOM} in which the problem of real-time tracking and reconstruction of an information source with the purpose of actuation is studied. These papers proposed semantics-empowered policies to achieve significant reduction in both the real-time reconstruction and the cost of actuation errors. In this work, we introduce a new state-aware sampling and transmission policy, and we evaluate its performance in terms of a set of semantics-aware metrics that capture the significance of information and various characteristics of the system's performance. Our key contributions are summarized as follows:
	\begin{enumerate}
		\item We propose a \emph{state-aware randomized stationary} sampling and transmission policy, in which we consider different sampling and success probabilities for different states of the information source. This becomes relevant to scenarios where the states encode commands for actuation or other potential tasks, where different actions have different importance, thus, is important to allow for different sampling frequencies.
		\item We analyze the performance of the proposed strategy in terms of time-averaged reconstruction error, cost of actuation error, and consecutive error metrics, and we compare it with previous proposed joint sampling and transmission policies \cite{pappas2021goal,salimnejad2023ICC,salimnejad2023TCOM}. 
		\item We define a new timing-aware error metric, namely \emph{importance-aware consecutive error} metric, which jointly captures both timing- and importance-related aspects of errors. Specifically, this metric measures the impact on the performance when the system remains in a specific erroneous state for several consecutive time slots. 
		\item We solve the optimization problem of minimizing the average cost of actuation error subject to a time-averaged sampling cost constraint, as a means to reveal when and under which conditions the proposed state-aware randomized stationary policy outperforms state-of-the-art alternatives.
	\end{enumerate}

	\section{System Model}
	\label{system model}
	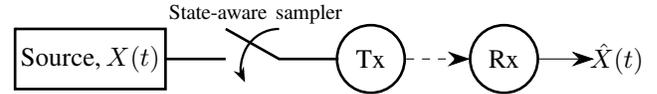
\begin{figure}[!t]
		\centering
		\begin{tikzpicture}
			\draw [line width=0.35mm](0,-3.5)--(2,-3.5)--(2,-4.3)--(0,-4.3)--(0,-3.485);
			\node(a) at (1,-3.9)  {\normalsize Source,\! $X(t)$};
			\draw [line width=0.35mm](2,-3.9)--(2.8,-3.9);
			\draw [line width=0.35mm](2.8,-3.5)--(3.51,-3.9);
			\draw [line width=0.35mm](3.5,-3.9)--(4.3,-3.9);
			
			\draw[line width=0.3mm,-{Stealth[length=2mm]}] (3.51,-3.5) arc (90:205:5mm) ;
			\draw [line width=0.35mm](4.73,-3.9) circle (0.45cm);
			\node(b) at (4.73,-3.9)  {\normalsize$\text{Tx}$};
			\draw [-{Stealth[length=3mm, width=2mm]},dashed](5.2,-3.9) -- (6.05,-3.9);
			\draw [line width=0.35mm](6.5,-3.9) circle (0.45cm);
			\node(c) at (6.5,-3.9)  {\normalsize$\text{Rx}$};
			\draw [-{Stealth[length=3mm, width=2mm]}](6.96,-3.9) -- (7.7,-3.9);
			\node(d) at (8,-3.9)  {\normalsize$\hat{X}(t)$};
			\node(e) at (3.2,-3.3)  {\footnotesize$\text{State-aware sampler}$};
		\end{tikzpicture}
		\vspace*{1ex}
		\caption{Real-time remote tracking of an information source over a wireless channel.}
		\label{system_model_fig}
	\end{figure}
	\par
\begin{figure}[!t]
		\centering
		\scriptsize
		\begin{tikzpicture}[start chain=going left,->,>=latex,node distance=2.5cm]
			\node[state, on chain]                 (2){$1$};
			\node[state, on chain]                 (1){$0$};
			\draw[>=latex]
			(1)   edge[loop above] node {$1-p$} ()
			(1) edge  [bend left] node[above] {$p$}(2)
			(2)   edge[loop above] node {$1-q$} ()
			(2) edge  [bend left] node[above] {$q$}(1)
			;
		\end{tikzpicture}
		\vspace*{1ex}
		\caption{DTMC describing the evolution of the information source $X(t)$.}
		\label{InformationSource}
	\end{figure}
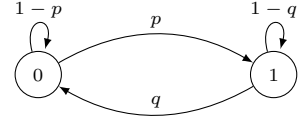

 We consider a time slotted communication system in which a sampler performs sampling of an information source $X(t)$ at time slot $t$, after which the transmitter sends the sample to the receiver over a wireless channel, as shown in Fig. \ref{system_model_fig}. The remote receiver operates as an actuator and performs actions based on the reconstructed state of the information source. We model the information source as a two-state discrete time Markov chain (DTMC) $\{X(t), t \in \mathbb{N}\}$, depicted in Fig. \ref{InformationSource}. Therein, the self-transition probability and the probability of transition to another state at time slot $t+1$ are defined as follows
	\begin{align}
		\label{TransProb_InfoSource}
		\mathrm{Pr}\big[X(t+1)=i\big|X(t)=j\big] =
		\begin{cases}
			1-p, &i=0,j=0\\
			q, &i=0,j=1\\
			p, &i=1,j=0\\
			1-q, &i=1,j=1.
		\end{cases}
	\end{align}
	
	\par In this paper, we consider different sampling and transmission actions for the states of the information source. We denote the action of sampling at time slot $t$ when the information source is at state $i$ $(i = 0,1)$ by $\alpha^{\text{s}}_{i}(t)$, where $\alpha^{\text{s}}_{i}(t)=1$ if the source at state $i$ is sampled and $\alpha^{\text{s}}_{i}(t)=0$ otherwise. Furthermore, when $\alpha^{\text{s}}_{i}(t)=1$, the action of transmitting the sample is denoted by $\alpha^{\text{tx}}_{i}(t)$, where $\alpha^{\text{tx}}_{i}(t)=1$ if the sample is transmitted, otherwise the transmitter remains idle, $\alpha^{\text{tx}}_{i}(t)=0$. At time slot $t$, the receiver constructs an estimate of the process $X(t)$, denoted by $\hat{X}(t)$ based on successfully received samples. The channel state $h_{i}(t)$ is equal to $1$ if the information source at state $i$ is sampled and successfully decoded by the receiver, and $0$ otherwise. We define the success probability when the information source at state $i$ is sampled and transmitted, as $p_{\text{s}_{i}} = \mathrm{Pr}\big[h_{i}(t)=1\big]$. Note that allowing for different success probabilities can have interesting connections with performing simple state-aware power control. Successful/failed transmissions are declared to the transmitter using acknowledgment (ACK)/negative-ACK packets, which are assumed to be delivered instantaneously and error free to the transmitter\footnote{Actually, only the semantics-aware policy requires an ACK/NACK feedback channel.}. Therefore, the transmitter has perfect knowledge of the reconstructed source state at time slot $t$, i.e., $\hat{X}(t)$ \footnote{In this paper, since we do not consider any decoder/estimation policy, the current state at the transmitter is the latest received update.}. We also assume that a sample is discarded when its transmission fails.
	
	\section{Sampling and Transmission Policies}
	\par We propose a new sampling and transmission policy, coined \emph{state-aware randomized stationary}, in which the generation of a sample is triggered in a probabilistic manner at each time slot. Specifically, we introduce a scheme that allows for assigning different sampling probabilities for different states, adjusting the sampling frequency depending on the importance of the state. Consider, for example, the scenario where each state is a command for a remote agent that requires to be executed, and different commands are of different importance or criticality. We assume that $p_{\alpha^{\text{s}}_{i}}$ is the probability of joint sampling and transmission actions when the source is at the state $i$. Therefore, we define $p_{\alpha^{\text{s}}_{i}}$ as follows
		\begin{align}
			\mathrm{Pr}\Big[\alpha^{\text{s}}_{i}(t+1)=1, \alpha^{\text{tx}}_{i}(t+1)=1\Big] &= p_{\alpha^{\text{s}}_{i}}.
		\end{align}
		The probability that the source at the state $i$ is not sampled at time slot $t+1$ is $\mathrm{Pr}\Big[\alpha^{\text{s}}_{i}(t+1) = 0\Big]=1-p_{\alpha^{\text{s}}_{i}}$. 
	 In addition, for comparison, we adopt three relevant policies proposed in \cite{pappas2021goal,salimnejad2023TCOM}. Below we provide a short description of them.
	\begin{enumerate}
		\item \emph{Uniform:} sampling is conducted periodically every $d$ time slots, independently of the evolution of the source $X(t)$. Therefore, the sampling time sequences are $\{t_{k} = k d, k\geqslant 1\}$. While this policy is simple and easy to implement, several state transitions can be missed during the time interval between two performed samples.
		\item \emph{Change-aware:} a new sample is generated when a change in the state of the source $X(t)$ is observed between two consecutive time slots without considering whether the system is in sync or not.
		\item \emph{Semantics-aware:} when the system is in a sync state, i.e., $X(t) = \hat{X}(t)$, a sample is generated if a change in the source state is observed at the next time slot, i.e., $X(t+1) \neq X(t)$. However, when the system is in an erroneous state, i.e., $X(t) \neq \hat{X}(t)$, a sample is generated if the source state at the next time slot is not equal to the state of the reconstructed source at time slot $t$, i.e., $X(t+1) \neq \hat{X}(t)$.
	\end{enumerate}
	
	\section{Preliminary Performance Analysis}
	\par In this section, we analyze the performance of state-aware randomized stationary policy in terms of time-averaged reconstruction error and average cost of actuation error. 
	
	\subsection{Real-time Reconstruction Error}
	\label{RealTime}
	\par The real-time reconstruction error captures the discrepancy between the original source $X(t)$ and the reconstructed source $\hat{X}(t)$ at time slot $t$, i.e.,
	\begin{align}
		\label{reconstructed_error}
		E(t) = \left|X(t)- \hat{X}(t)\right|,
	\end{align}
	where at time slot $t$, $E(t) = 0$ denotes the system is in the sync state, while the erroneous state of the system is denoted by $E(t)\neq 0$. The time-averaged reconstruction error or the probability that the system is in an erroneous state, $P_{E}$, for an observation interval $[1, T]$ with $T$ being a large positive number, is defined as \cite{pappas2021goal,salimnejad2023TCOM}
	\begin{align}
		\label{timeAvg_Error}
		P_{E}  \!=\!\! \lim_{T\to\infty} \frac{1}{T}\!\sum_{t = 1}^{T} \mathbbm{1}\left(E(t)\!\neq\! 0\right) \!=\!\! \lim_{T\to\infty} \frac{1}{T}\!\sum_{t = 1}^{T} \mathbbm{1}\!\left(X(t)\!\neq\! \hat{X}(t)\right),
	\end{align}
	where $\mathbbm{1}(\cdot)$ is the indicator function. 
	
	For a two-state DTMC information source, $P_{E}$ in \eqref{timeAvg_Error} is given by
	\begin{align}
		\label{PE}
		P_{E}&= \mathrm{Pr}[X(t)=0, \hat{X}(t)=1]+\mathrm{Pr}[X(t)=1, \hat{X}(t)=0] \notag\\
		&= \pi_{0,1}+\pi_{1,0},
	\end{align}
	Note that $\pi_{0,1}$ and $\pi_{1,0}$ are the probabilities obtained from the stationary distribution of the two-dimensional DTMC describing the joint status of the system regarding the current state at the original source, i.e., $\big(X(t),\hat{X}(t)\big)$.
	To derive $\pi_{i,j}, (i,j)\in \{0,1\}$, we assume that when the sampler performs sampling, the transmitter sends the sample in the form of packets during the same time slot.  
	\begin{lemma}
		\label{lemma_PI_ij_DTMCmodel}
		For a two-state DTMC information source, the stationary distribution $\pi_{i,j}$ for the  state-aware randomized stationary  policy is given by\footnote{This work can be extended to more than two-state DTMC information source. In Appendix~\ref{3States}, we provide an example for the three-state DTMC information source.}
		\begin{subequations}
			\begin{align}
				\pi_{0,0}&=\frac{q p_{\alpha^{\text{s}}_{0}} p_{{\text{s}}_{0}} \big[q+(1-q)p_{\alpha^{\text{s}}_{1}} p_{{\text{s}}_{1}}\big]}{(p+ q) \Phi\big(p_{\alpha^{\text{s}}_{0}},p_{\alpha^{\text{s}}_{1}}\big)}\label{pi00}\\
				\pi_{0,1}&=\frac{pq p_{\alpha^{\text{s}}_{1}} p_{{\text{s}}_{1}} \big(1-p_{\alpha^{\text{s}}_{0}} p_{{\text{s}}_{0}}\big)}{(p+ q) \Phi\big(p_{\alpha^{\text{s}}_{0}},p_{\alpha^{\text{s}}_{1}}\big)}\label{pi01}\\
				\pi_{1,0}&=\frac{pq p_{\alpha^{\text{s}}_{0}} p_{{\text{s}}_{0}} \big(1-p_{\alpha^{\text{s}}_{1}} p_{{\text{s}}_{1}}\big)}{(p+ q) \Phi\big(p_{\alpha^{\text{s}}_{0}},p_{\alpha^{\text{s}}_{1}}\big)}\label{pi10}\\
				\pi_{1,1}&=\frac{p p_{\alpha^{\text{s}}_{1}} p_{{\text{s}}_{1}} \big[p+(1-p)p_{\alpha^{\text{s}}_{0}} p_{{\text{s}}_{0}}\big]}{(p+ q) \Phi\big(p_{\alpha^{\text{s}}_{0}},p_{\alpha^{\text{s}}_{1}}\big)}\label{pi11},
			\end{align}
		\end{subequations}
		where 
		\begin{align}
			\label{Dnum}
			\!\Phi\big(p_{\alpha^{\text{s}}_{0}},p_{\alpha^{\text{s}}_{1}}\big) \!=\! p p_{\alpha^{\text{s}}_{1}} p_{{\text{s}}_{1}} \big(1 \!-\! p_{\alpha^{\text{s}}_{0}} p_{{\text{s}}_{0}}\big) \!+\! p_{\alpha^{\text{s}}_{0}} p_{{\text{s}}_{0}}\big(q\!+\!(1\!-\!q)p_{\alpha^{\text{s}}_{1}} p_{{\text{s}}_{1}}\big)\!.
		\end{align}
		Proof. See Appendix~\ref{Appendixlemma1}.
	\end{lemma}
	\par Using eqs. \eqref{pi01} and \eqref{pi10}, the time-averaged reconstruction error in \eqref{timeAvg_Error} can be calculated as
	\begin{align}
		\label{TimeAveraged_ReconstError}
		P_{E}=\pi_{0,1}+\pi_{1,0}
		=\frac{pq\Big[p_{\alpha^{\text{s}}_{1}} p_{{\text{s}}_{1}} + p_{\alpha^{\text{s}}_{0}} p_{{\text{s}}_{0}} \big(1 - 2 p_{\alpha^{\text{s}}_{1}} p_{{\text{s}}_{1}}\big)\Big]}{(p+ q) \Phi\big(p_{\alpha^{\text{s}}_{0}},p_{\alpha^{\text{s}}_{1}}\big)},
	\end{align}
	where $ \Phi\big(p_{\alpha^{\text{s}}_{0}},p_{\alpha^{\text{s}}_{1}}\big)$ is given in \eqref{Dnum}.
	\subsection{Cost of Actuation Error}
	\label{cost_Act_error}
	\par This metric captures the significance of the error at the receiver side and considers different cost or penalties for different erroneous actions. To study the cost of actuation error we define $C_{i,j}$ the cost of error when the current state of the source is $i$, and the reconstructed source is in state $j\neq i$. It is assumed that $C_{i,j}$ does not change over time. Now, using $C_{i,j}$, the average cost of actuation error for a two-state DTMC can be calculated as follows
	\begin{align}
		\label{Averaged_CostActuationError}
		P^{C}_{E}=C_{0,1}\pi_{0,1}+C_{1,0}\pi_{1,0},
	\end{align}
	where using eqs. \eqref{pi01} and \eqref{pi10}, we can write \eqref{Averaged_CostActuationError} as
	\begin{multline}
		\label{Avg_cost_error}
		P^{C}_{E}\!=\!\\
		\frac{pq\Big[C_{0,1} p_{\alpha^{\text{s}}_{1}} p_{{\text{s}}_{1}}\big(1\!-\! p_{\alpha^{\text{s}}_{0}} p_{{\text{s}}_{0}}\big)\!+\!C_{1,0} p_{\alpha^{\text{s}}_{0}} p_{{\text{s}}_{0}}\big(1\!-\! p_{\alpha^{\text{s}}_{1}} p_{{\text{s}}_{1}}\big)\Big]}{(p+ q)\Big[p p_{\alpha^{\text{s}}_{1}} p_{{\text{s}}_{1}} \big(1\!-\!p_{\alpha^{\text{s}}_{0}} p_{{\text{s}}_{0}}\big) \!+\! p_{\alpha^{\text{s}}_{0}} p_{{\text{s}}_{0}}\big(q\!+\!(1\!-\!q)p_{\alpha^{\text{s}}_{1}} p_{{\text{s}}_{1}}\big)\Big]}.
	\end{multline}
	\begin{remark}
		\label{remark_RS_SA_compare}
		Using \eqref{pij} and \eqref{pij_SA}, we can prove that when $\max\{0,T_{1}\}\leqslant p_{\alpha^{\text{s}}_{1}} \leqslant 1$, the state-aware randomized stationary  policy has lower average cost of actuation error as compared to the semantics-aware policy for $\max\{0,T_{2}\}\leqslant p_{\alpha^{\text{s}}_{0}} \leqslant 1$, where $T_{1}$ and $T_{2}$ are given by
		\begin{align}
			\label{T1_T2}
			T_{1} &= \frac{pC_{1,0}+C_{1,0}p_{{\text{s}}_{0}}-pC_{1,0}p_{{\text{s}}_{0}}-qC_{0,1}(1-p_{{\text{s}}_{0}})}{C_{1,0}(1-p)p_{{\text{s}}_{0}}+pC_{1,0}p_{{\text{s}}_{1}}+C_{0,1}(1-q)p_{{\text{s}}_{1}}+qC_{0,1}p_{{\text{s}}_{0}}}\notag\\
			T_{2} &= {p_{\alpha^{\text{s}}_{1}}\big[C_{0,1}\big(q+(1-q)p_{{\text{s}}_{1}}\big)-pC_{1,0}(1-p_{{\text{s}}_{1}})\big]}\notag\\
			&\!\times\!\Big[p_{\alpha^{\text{s}}_{1}}\big(C_{1,0}p_{{\text{s}}_{0}}(1\!-\!p)\!+\!pC_{1,0}p_{{\text{s}}_{1}}\!+\!C_{0,1}p_{{\text{s}}_{1}}(1\!-\!q)\!\notag\\
			&+\!qC_{0,1}p_{{\text{s}}_{0}}\big)
			\!-\!pC_{1,0}\!-\!C_{1,0}p_{{\text{s}}_{0}}\!+\!pC_{1,0}p_{{\text{s}}_{0}}\!+\!qC_{0,1}(1-p_{{\text{s}}_{0}})\Big]^{-1}\!.
		\end{align}
		Also, when $0\leqslant p_{\alpha^{\text{s}}_{1}} \leqslant \min\{0,T_{1}\}$ and $0\leqslant p_{\alpha^{\text{s}}_{0}} \leqslant \min\{0,T_{2}\}$, the state-aware randomized stationary  policy has lower average cost of actuation error in comparison with the semantics-aware policy.
	\end{remark}
	\begin{remark}
		\label{remark_RS_pq}
		We can analytically prove that for $\frac{q^{2}p_{\alpha^{\text{s}}_{0}} p_{{\text{s}}_{0}}}{p_{{\text{s}}_{1}}\big[1-p_{\alpha^{\text{s}}_{0}} p_{{\text{s}}_{0}}\big(1+q(1-q)\big)\big]}\leqslant p_{\alpha^{\text{s}}_{1}}\leqslant 1$, the time-averaged reconstruction error and the average cost of actuation error are decreasing with $p$, when $\sqrt{\frac{qp_{\alpha^{\text{s}}_{0}} p_{{\text{s}}_{0}}\big(q+(1-q)p_{\alpha^{\text{s}}_{1}} p_{{\text{s}}_{1}}\big)}{p_{\alpha^{\text{s}}_{1}} p_{{\text{s}}_{1}}\big(1-p_{\alpha^{\text{s}}_{0}} p_{{\text{s}}_{0}}\big)}}<p\leqslant 1$. Furthermore, when $\frac{p^{2}p_{\alpha^{\text{s}}_{1}} p_{{\text{s}}_{1}}}{p_{{\text{s}}_{0}}\big[1-p_{\alpha^{\text{s}}_{1}} p_{{\text{s}}_{1}}\big(1+p(1-p)\big)\big]}\leqslant p_{\alpha^{\text{s}}_{0}}\leqslant 1$, the time-averaged reconstruction error and the average cost of actuation error are decreasing with $q$, for $\sqrt{\frac{pp_{\alpha^{\text{s}}_{0}} p_{{\text{s}}_{0}}p_{\alpha^{\text{s}}_{1}} p_{{\text{s}}_{1}}+p^{2}p_{\alpha^{\text{s}}_{1}} p_{{\text{s}}_{1}}\big(1-p_{\alpha^{\text{s}}_{0}} p_{{\text{s}}_{0}}\big)}{p_{\alpha^{\text{s}}_{0}} p_{{\text{s}}_{0}}\big(1-p_{\alpha^{\text{s}}_{1}} p_{{\text{s}}_{1}}\big)}}<q\leqslant 1$.
	\end{remark}
	
	\section{Joint Timing and Importance Error Metrics}
	
	In this section we consider the impact of timing and importance of errors, and we propose an extension of the consecutive error metric, termed \emph{importance-aware consecutive errors}, which takes into account \textit{jointly both the timing and the importance aspects} of errors.
	
	\subsection{Consecutive Error Metric}
	\label{Cons_error}
	\par
\begin{figure}[!t]
	\centering
	\begin{tikzpicture}[start chain=going left,->,>=latex,node distance=2cm,on grid,auto]
		\footnotesize
		\node[on chain]                        (g) {$\cdots$};
		\node[state, on chain]                 (3) {$2$};
		\node[state, on chain]                 (2) {$1$};
		\node[state, on chain]                 (1) {$0$};
		\draw[>=latex]
		(1)   edge[loop above] node {$1-P_{0,1}$}   (1)
		(1) edge  [bend left=30] node {$P_{0,1}$} (2)
		(2) edge  [bend left=30] node {$P_{1,2}$} (3)
		(2) edge  [bend left=30] node[above] {$1\!-\!P_{1,2}$} (1)
		(3) edge  [bend left=40] node {$1-P_{2,3}$} (1)
		(3) edge  [bend left=30] node {$P_{2,3}$} (g)
		(g) edge  [bend left=50] node {$1-P_{n,n+1}$} (1)
		;
	\end{tikzpicture}
	\caption{DTMC describing the state of the consecutive error.}
	\label{Cons_error_DTMC}
\end{figure}
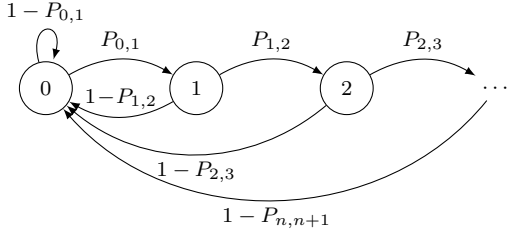

 The \emph{consecutive error} metric, first introduced in \cite{salimnejad2023TCOM}, quantifies the number of consecutive time slots during which the system is in an erroneous state \footnote{A similar metric was defined first in \cite{StamatakisGCW19} and then in \cite{AoIITON20}.}. This metric can be described by a DTMC as depicted in Fig. \ref{Cons_error_DTMC}. At time slot $t$, $C_{E}(t)=0$ denotes the synced state, whereas $C_{E}(t) = 1\leqslant i\leqslant n-1$ denotes the number of consecutive time slots for which the system is in an erroneous state. 
Furthermore, the transition probability $P_{i,i+1}$ is defined as $P_{i,i+1}=\mathrm{Pr}\big[C_E(t+1)=i+1\big|C_E(t)=i\big]$. For the state-aware randomized stationary policy, this transition probability is given by 
\begin{align}
	\label{P01_P11_RS}
	P_{i,i+1}&= \mathrm{Pr}\big[C_{E}(t+1)=i+1\big|C_{E}(t)=i\big]\notag\\
	&=\frac{\mathrm{Pr}\big[C_{E}(t)=i+1\big]}{\mathrm{Pr}\big[C_{E}(t)=i\big]}, \hspace{0.2cm}\forall i\geqslant 0,
\end{align}
where $\mathrm{Pr}[C_{E}(t)=i]$ for $i=0$ is equal to   $\mathrm{Pr}\big[C_{E}(t)=0\big]=\pi_{0,0}+\pi_{1,1}$, and for $i\geqslant 1$, it is calculated as (see Appendix \ref{Proof_P01_P11_RS})  
\begin{align}
	\label{Pr_CE_i}
	&\mathrm{Pr}\big[C_{E}(t)=i\big] \notag\\
	&\!\!=\!p(1\!-\!q)^{i-1}\big(1\!-\!p_{\alpha^{\text{s}}_{1}} p_{{\text{s}}_{1}}\big)^{i}\!\pi_{0,0}\!+\!q(1\!-\!p)^{i-1}\big(1\!-\!p_{\alpha^{\text{s}}_{0}} p_{{\text{s}}_{0}}\big)^{i}\!\pi_{1,1},
\end{align}
where $\pi_{i,j}, \forall i,j\in\{0,1\}$ was given in Lemma \ref{lemma_PI_ij_DTMCmodel}. Now, using $\mathrm{Pr}\big[C_{E}(t)=0\big]=\pi_{0,0}+\pi_{1,1}$ and \eqref{Pr_CE_i}, the transition probability given in \eqref{P01_P11_RS} can be written as
\begin{subequations}
	\label{Trans_Prob_ij}
	\begin{align}
		&P_{0,1}=\frac{p\big(1-p_{\alpha^{\text{s}}_{1}} p_{{\text{s}}_{1}}\big)\pi_{0,0}+q\big(1-p_{\alpha^{\text{s}}_{0}} p_{{\text{s}}_{0}}\big)\pi_{1,1}}{\pi_{0,0}+\pi_{1,1}},\\
		&P_{i,i+1}\notag\\
		&=\frac{\!p(1\!-\!q)^{i}\big(1\!-\!p_{\alpha^{\text{s}}_{1}} p_{{\text{s}}_{1}}\big)^{i+1}\!\pi_{0,0}\!+\!q(1\!-\!p)^{i}\big(1\!-\!p_{\alpha^{\text{s}}_{0}} p_{{\text{s}}_{0}}\big)^{i+1}\!\pi_{1,1}}{\!p(1\!-\!q)^{i-1}\big(1\!-\!p_{\alpha^{\text{s}}_{1}} p_{{\text{s}}_{1}}\big)^{i}\!\pi_{0,0}\!+\!q(1\!-\!p)^{i-1}\big(1\!-\!p_{\alpha^{\text{s}}_{0}} p_{{\text{s}}_{0}}\big)^{i}\!\pi_{1,1}}.
	\end{align}
\end{subequations}

Using \eqref{Pr_CE_i}, we can calculate the average consecutive error $\Bar{C}_{E}$ as
\begin{align}
	\label{Avg_Cons_error}
	\bar{C}_{E} &= \sum_{x = 1}^{\infty} x\mathrm{Pr}\big[C_{E}(t)=x\big]\notag\\
	&= \frac{p\big(1-p_{\alpha^{\text{s}}_{1}} p_{{\text{s}}_{1}}\big)\pi_{0,0}}{\big(q+(1-q)p_{\alpha^{\text{s}}_{1}} p_{{\text{s}}_{1}}\big)^2}+\frac{q\big(1-p_{\alpha^{\text{s}}_{0}} p_{{\text{s}}_{0}}\big)\pi_{1,1}}{\big(p+(1-p)p_{\alpha^{\text{s}}_{0}} p_{{\text{s}}_{0}}\big)^2}.
\end{align}
Note, that the convergence conditions for the previous expression are $\left|(1-p)\big(1-p_{\alpha^{\text{s}}_{0}} p_{{\text{s}}_{0}}\big)\right|<1$ and $\left|(1-q)\big(1-p_{\alpha^{\text{s}}_{1}} p_{{\text{s}}_{1}}\big)\right|<1$.

In the following, we define the \textit{consecutive error violation probability} metric as the percentage of time during which the system remains in an erroneous state for more than $n$ consecutive time slots. Therefore, using the expression given in \eqref{Pr_CE_i} and Lemma \ref{lemma_PI_ij_DTMCmodel}, we can write
\begin{align}
	\label{ViolationProb}
	\mathrm{Pr}\big[C_{E}(t)>n\big]&=\sum_{x = n+1}^{\infty}\mathrm{Pr}\big[C_{E}(t)=x\big]\notag\\
	&= \frac{pqp_{\alpha^{\text{s}}_{0}} p_{{\text{s}}_{0}}\big[(1-q)\big(1-p_{\alpha^{\text{s}}_{1}} p_{{\text{s}}_{1}}\big)\big]^{n+1}}{(1-q)(p+q)\Phi\big(p_{\alpha^{\text{s}}_{0}},p_{\alpha^{\text{s}}_{1}}\big)}\notag\\
	&+\frac{pqp_{\alpha^{\text{s}}_{1}} p_{{\text{s}}_{1}}\big[(1-p)\big(1-p_{\alpha^{\text{s}}_{0}} p_{{\text{s}}_{0}}\big)\big]^{n+1}}{(1-p)(p+q)\Phi\big(p_{\alpha^{\text{s}}_{0}},p_{\alpha^{\text{s}}_{1}}\big)},
\end{align}
where $n$ in \eqref{ViolationProb} $n$ is finite.

\subsection{Importance-Aware Consecutive Errors}
\label{ImpotanceAware}
\par In this section, we introduce a new timing-aware error metric as a means to capture the significance of a particular erroneous action at the receiver side. For that, we propose the \emph{importance-aware consecutive error} metric, which is defined as the number of consecutive time slots that the system is in an particular erroneous state. Here, we assume that the system is in an erroneous action when the state of the source is $1$ $\big(X(t) = 1\big)$, and the reconstructed source is in state $0$ $\big(\hat{X}(t) = 0\big)$. Now, let $S(t)\neq 0$ denote that the system is in the mentioned erroneous state at time slot $t$, while the synced state of the system is denoted by $S(t)=0$. We also define $C_{S}(t)$ as the consecutive error at time slot $t$ when the system is in the mentioned particular erroneous state. Now, we can define the state evolution of consecutive error as follows
\begin{align}
	\label{CSt}
	C_{S}(t+1) = 
	\begin{cases}
		C_{S}(t)+1,\hspace{0.2cm}& X(t+1)=1,\hat{X}(t+1)=0\\
		0,\hspace{0.2cm} &\text{otherwise}.
	\end{cases}
\end{align}
Using \eqref{CSt}, we define the transition probability of $C_{S}(t)$ as 
\begin{align}
	\label{Pij_S}
	P^{S}_{i,i+1} &= \mathrm{Pr}\big[C_{S}(t+1)=i+1\big|C_{S}(t)=i\big]\notag\\
	&=\frac{\mathrm{Pr}\big[C_{S}(t+1)=i+1\big]}{\mathrm{Pr}\big[C_{S}(t+1)=i\big]}.
\end{align}
Now, using the similar procedure presented in Section \ref{Cons_error}, one can obtain $\mathrm{Pr}\big[C_{S}(t)=i\big]$ as follows
\begin{align}
	\label{Pr_CS_i}
	&\mathrm{Pr}\big[C_{S}(t)=i\big] =
	\begin{cases}
		1-\pi_{1,0}, \hspace{0.2cm} i=0,\\
		p(1-q)^{i-1}\big(1-p_{\alpha^{\text{s}}_{1}} p_{{\text{s}}_{1}}\big)^{i}\pi_{0,0}, \hspace{0.2cm} i\geqslant 1.
	\end{cases}
\end{align}
where $\pi_{0,0}$ and $\pi_{1,0}$ are given in Lemma {\ref{lemma_PI_ij_DTMCmodel}}. Now, using \eqref{Pr_CS_i} we can write \eqref{Pij_S} as
\begin{align}
	P^{S}_{i,i+1} =
	\begin{cases}
		\displaystyle \frac{p\big(1-p_{\alpha^{\text{s}}_{1}} p_{{\text{s}}_{1}}\big)\pi_{0,0}}{1-\pi_{1,0}}, \hspace{0.2cm} &i=0\\
		(1-q)\big(1-p_{\alpha^{\text{s}}_{1}} p_{{\text{s}}_{1}}\big), \hspace{0.2cm} &i\geqslant 1.
	\end{cases}
\end{align}
\begin{remark}
	The transition probability $P^{S}_{i,0} \hspace{0.2cm}(i\geqslant 0)$, is defined as $P^{S}_{i,0} = 1-P^{S}_{i,i+1}$.
\end{remark}

Now, using \eqref{Pr_CS_i} and Lemma \ref{lemma_PI_ij_DTMCmodel}, the average importance-aware consecutive error, $\Bar{C}_{S}$, can be obtained as
\begin{align}
	\Bar{C}_{S} &= \sum_{x=1}^{\infty}x\mathrm{Pr}\big[C_{S}(t)=x\big] \notag\\&= \frac{pqp_{\alpha^{\text{s}}_{0}} p_{{\text{s}}_{0}}\big(1-p_{\alpha^{\text{s}}_{1}} p_{{\text{s}}_{1}}\big)}{(p+q)\big(q+(1-q)p_{\alpha^{\text{s}}_{1}} p_{{\text{s}}_{1}}\big)\Phi\big(p_{\alpha^{\text{s}}_{0}},p_{\alpha^{\text{s}}_{1}}\big)},p_{\alpha^{\text{s}}_{0}},p_{\alpha^{\text{s}}_{1}}\neq 0,
\end{align}
where $ \Phi\big(p_{\alpha^{\text{s}}_{0}},p_{\alpha^{\text{s}}_{1}}\big)$ is given in \eqref{Dnum}.

\begin{remark}
	We would like to emphasize that since the metric of importance-aware consecutive errors considers exclusively the scenario of a specific error, this metric needs to be studied in combination with another error metric, for instance, considering a constrained optimization problem or combining the metrics to form \textit{error vectors}. This will become prominent in the numerical results section.
\end{remark}

\section{Optimization Problem}
\label{optimizationproblem}
\par In this section, our objective is to find an optimal state-aware randomized stationary sampling policy, which minimizes the average cost of actuation error subject to a time-averaged sampling cost constraint. Here, we assume that each attempted sampling has a sampling cost $\delta$, and that the time-averaged sampling cost cannot exceed a certain threshold $\delta_{\text{max}}$. Therefore, the optimization problem is formulated as
\begin{subequations}
	\label{Optimization_problem1}
	\begin{align}
		&\underset{p_{\alpha^{\text{s}}_{0}},  p_{\alpha^{\text{s}}_{1}}}{\text{minimize}}\hspace{0.3cm}\hspace{0.3cm} P^{C}_{E}\\
		&\text{subject to}\hspace{0.3cm} \lim_{T \to \infty}\frac{1}{T}\sum_{t=1}^{T}\delta \mathbbm{1}\{\alpha^{\text{s}}_{t}=1\} \leqslant\delta_{\text{max}},\label{Optimization_prob1_constraint}
	\end{align}
\end{subequations}
where the constraint given in \eqref{Optimization_prob1_constraint} can be written as
\begin{align}
	\label{constraint1}
	\lim_{T \to \infty}\!\frac{1}{T}\!\sum_{t=1}^{T}\!\delta \mathbbm{1}\{\alpha^{\text{s}}_{t}\!=\!1\}\!&=\!\delta \mathrm{Pr}[X(t)\!=\!0]p_{\alpha^{\text{s}}_{0}}\!+\!\delta\mathrm{Pr}[X(t)\!=\!1]p_{\alpha^{\text{s}}_{1}}\notag\\
	&=\delta\frac{qp_{\alpha^{\text{s}}_{0}}}{p+q}+\delta\frac{pp_{\alpha^{\text{s}}_{1}}}{p+q}.
\end{align}
Now, using eqs. \eqref{Averaged_CostActuationError} and \eqref{constraint1}, the optimization problem can be simplified as
\begin{subequations}
	\label{Optimization_problem2}
	\begin{align}
		&\underset{p_{\alpha^{\text{s}}_{0}},  p_{\alpha^{\text{s}}_{1}}}{\text{minimize}}\hspace{0.3cm}\hspace{0.3cm} \frac{pq\Psi \big(p_{\alpha^{\text{s}}_{0}},p_{\alpha^{\text{s}}_{1}}\big)}{(p+ q)\Phi \big(p_{\alpha^{\text{s}}_{0}},p_{\alpha^{\text{s}}_{1}}\big)}\label{Optimization_prob2_objfun}\\
		&\text{subject to}\hspace{0.3cm} qp_{\alpha^{\text{s}}_{0}}+pp_{\alpha^{\text{s}}_{1}} \leqslant\eta(p+q),\label{Optimization_prob2_constraint}
	\end{align}
\end{subequations}
where $\eta=\delta_{\text{max}}/\delta$, $\Psi \big(p_{\alpha^{\text{s}}_{0}},p_{\alpha^{\text{s}}_{1}}\big) = C_{0,1} p_{\alpha^{\text{s}}_{1}} p_{{\text{s}}_{1}}\big(1- p_{\alpha^{\text{s}}_{0}} p_{{\text{s}}_{0}}\big)+C_{1,0} p_{\alpha^{\text{s}}_{0}} p_{{\text{s}}_{0}}\big(1- p_{\alpha^{\text{s}}_{1}} p_{{\text{s}}_{1}}\big)$, and $\Phi \big(p_{\alpha^{\text{s}}_{0}},p_{\alpha^{\text{s}}_{1}}\big)$ is given by \eqref{Dnum}. 

To solve this optimization problem, we first note that the objective function in \eqref{Optimization_prob2_objfun} is decreasing with $p_{\alpha^{\text{s}}_{0}}$, i.e., $\frac{\partial P^{C}_{E}}{\partial p_{\alpha^{\text{s}}_{0}}}<0$, when
\begin{align}
	\label{cond1}
	p_{\alpha^{\text{s}}_{1}}\geqslant \frac{pC_{1,0}-qC_{0,1}}{p_{\text{s}_{1}}\big(pC_{1,0}+(1-q)C_{0,1}\big)}.
\end{align}
Also, the objective function is decreasing with $p_{\alpha^{\text{s}}_{1}}$ when
\begin{align}
	\label{cond2}
	p_{\alpha^{\text{s}}_{0}}\geqslant \frac{qC_{0,1}-pC_{1,0}}{p_{\text{s}_{0}}\big(qC_{0,1}+(1-p)C_{1,0}\big)}.
\end{align}
Based on \eqref{cond1} and \eqref{cond2}, we consider two cases: one with $pC_{1,0}\geqslant qC_{0,1}$ and one other with $pC_{1,0}< qC_{0,1}$.
	\begin{enumerate}
		\item \emph{When $pC_{1,0}\geqslant qC_{0,1}$}: in this case, we can always find a probability $p_{\alpha^{\text{s}}_{0}}\in [0,1]$ that satisfies the condition given in \eqref{cond2}. Therefore, since $p_{\alpha^{\text{s}}_{0}}\geqslant 0$, using \eqref{cond2}, the objective function has its minimum value when $p_{\alpha^{\text{s}}_{1}}$ is maximized. Now, using the constraint given in \eqref{Optimization_prob2_constraint}, the maximum value of $p_{\alpha^{\text{s}}_{1}}$ is 
		\begin{align}
			\label{pa1_optimal0}
			p_{\alpha^{\text{s}}_{1}}=\frac{\eta(p+q)-qp_{\alpha^{\text{s}}_{0}}}{p}.
		\end{align}
		Using \eqref{pa1_optimal0}, the optimization problem can be written as
		\begin{subequations}
			\label{Optimization_problem3}
			\begin{align}
				&\underset{p_{\alpha^{\text{s}}_{0}}}{\text{minimize}}\hspace{0.3cm}\hspace{0.3cm} \frac{F}{G}\label{Optimization_problem3_objfunc}\\
				&\text{subject to}\hspace{0.3cm}	p^{\text{LB}}_{\alpha^{\text{s}}_{0}}\leqslant p_{\alpha^{\text{s}}_{0}}\leqslant 	p^{\text{UB}}_{\alpha^{\text{s}}_{0}} ,\label{Optimization_prob3_constraint}
			\end{align}
		\end{subequations}
		where $F$, $G$, $p^{\text{LB}}_{\alpha^{\text{s}}_{0}}$, and $p^{\text{UB}}_{\alpha^{\text{s}}_{0}}$ are given by
		\begin{align}
			\label{phi_I}
			F&=A_{1}p_{\alpha^{\text{s}}_{0}}^{2}\!+\!A_{2}p_{\alpha^{\text{s}}_{0}}\!+\!A_{3},\hspace{0.1cm}
			G=B_{1}p_{\alpha^{\text{s}}_{0}}^{2}\!+\!B_{2}p_{\alpha^{\text{s}}_{0}}\!+\!B_{3}.\notag\\
			p^{\text{LB}}_{\alpha^{\text{s}}_{0}} &= \text{max}\left\{\!0,\frac{\eta(p\!+\!q)\!-\!p}{q}\!\right\}\!,
			p^{\text{UB}}_{\alpha^{\text{s}}_{0}} \!=\!\text{min}\left\{\!1,\frac{\eta(p+q)}{q}\!\right\}\!,
		\end{align}
		where $A_{i}$, and $B_{i}$ $\forall i\in\{1,2,3\}$, are given by
		\begin{align}
			\label{coeff}
			A_{1}&=pq^{2}p_{{\text{s}}_{0}}p_{{\text{s}}_{1}}\big(C_{0,1}+C_{1,0}\big)\notag\\
			A_{2}&=pq\Big[C_{1,0}pp_{{\text{s}}_{0}}\big(1-\eta p_{{\text{s}}_{1}}\big)-qC_{0,1}p_{{\text{s}}_{1}}-q\eta C_{1,0}p_{{\text{s}}_{0}}p_{{\text{s}}_{1}}\notag\\
			&\hspace{3.75cm}-\eta (p+q)C_{0,1}p_{{\text{s}}_{0}}p_{{\text{s}}_{1}}\Big]\notag\\
			A_{3}&=pq(p+q)\eta C_{0,1}p_{{\text{s}}_{1}}\notag\\
			B_{1}&=(p+q)\Big[pqp_{{\text{s}}_{0}}p_{{\text{s}}_{1}}-q(1-q)p_{{\text{s}}_{0}}p_{{\text{s}}_{1}}\Big]\notag\\
			B_{2}&=(p+q)\Big[pqp_{{\text{s}}_{0}}-pqp_{{\text{s}}_{1}}-\eta p(p+q)p_{{\text{s}}_{0}}p_{{\text{s}}_{1}}\notag\\
			&\hspace{2.48cm}+\eta (1-q)(p+q)p_{{\text{s}}_{0}}p_{{\text{s}}_{1}}\Big]\notag\\
			B_{3}&=\eta p p_{{\text{s}}_{1}}(p+q)^{2}.
		\end{align}
		To determine the value of $p_{\alpha^{\text{s}}_{0}}$ that minimizes the objective function in \eqref{Optimization_problem3_objfunc}, we need to calculate the critical points of the objective function within the interval $\big[	p^{\text{LB}}_{\alpha^{\text{s}}_{0}},	p^{\text{UB}}_{\alpha^{\text{s}}_{0}}\big]$. When $\big(2A_{1}B_{3}-2A_{3}B_{1}\big)^{2}\geqslant 4\big(A_{1}B_{2}-A_{2}B_{1}\big)\big(A_{2}B_{3}-A_{3}B_{2}\big)$, one can obtain the critical points of the objective function by taking the first derivative $\frac{\partial }{\partial p_{\alpha^{\text{s}}_{0}}}\big(\frac{F}{G}\big)=0$ as\footnote{When $\Delta$ in \eqref{delta} is negative, the optimal value of $p_{\alpha^{\text{s}}_{0}}$ that minimizes the objective function in \eqref{Optimization_problem3_objfunc} is equal to $p^{\text{LB}}_{\alpha^{\text{s}}_{0}}$ if $A_{1}B_{2}> A_{2}B_{1}$ and $p^{\text{UB}}_{\alpha^{\text{s}}_{0}}$ if $A_{1}B_{2}< A_{2}B_{1}$.}
		\begin{subequations}
			\label{CriticalPoints}
			\begin{align}
				p_{\alpha^{\text{s}}_{0}}&=\frac{2\big(A_{3}B_{1}-A_{1}B_{3}\big)\pm\sqrt{\Delta}}{2\big(A_{1}B_{2}-A_{2}B_{1}\big)},\hspace{0.2cm} A_{1}B_{2}\neq A_{2}B_{1}\\
				p_{\alpha^{\text{s}}_{0}}&=\frac{A_{3}B_{2}\!-\!A_{2}B_{3}}{2\big(A_{1}B_{3}\!-\!A_{3}B_{1}\big)},\hspace{0.05cm} A_{1}B_{2} \!=\! A_{2}B_{1},\hspace{0.1cm} A_{1}B_{3}\!\neq\! A_{3}B_{1},
			\end{align}
		\end{subequations}
		where $A_{i}$, and $B_{i}$ are given in \eqref{coeff} and $\Delta$ can be written as
		\begin{align}
			\label{delta}
			\!\!\Delta\!\!=\!\!
			\big(2A_{1}B_{3}\!\!-\!\!2A_{3}B_{1}\big)^{2}\!\!-\!\! 4\big(A_{1}B_{2}\!-\!A_{2}B_{1}\big)\big(A_{2}B_{3}\!\!-\!\!A_{3}B_{2}\big).
		\end{align}
		Note that we consider only the value of $p_{\alpha^{\text{s}}_{0}}$ in \eqref{CriticalPoints} within the interval $\big[	p^{\text{LB}}_{\alpha^{\text{s}}_{0}},	p^{\text{UB}}_{\alpha^{\text{s}}_{0}}\big]$. Now, we evaluate the objective function at the critical points, as well as at points $p^{\text{LB}}_{\alpha^{\text{s}}_{0}}$ and $p^{\text{UB}}_{\alpha^{\text{s}}_{0}}$. The minimum value of  the objective function within the given interval corresponds to the smallest value. After determining the value of $p_{\alpha^{\text{s}}_{0}}$ that minimizes the objective function, we can calculate $p_{\alpha^{\text{s}}_{1}}$ by utilizing the expression given in \eqref{pa1_optimal0}. We note that the values of $p_{\alpha^{\text{s}}_{0}}$ and $p_{\alpha^{\text{s}}_{1}}$ obtained by solving the optimization problem in \eqref{Optimization_problem3}, are the optimal values of the sampling probabilities when $p_{\alpha^{\text{s}}_{1}}\geqslant \frac{pC_{1,0}-qC_{0,1}}{p_{\text{s}_{1}}\big(pC_{1,0}+(1-q)C_{0,1}\big)}$. Otherwise, the optimal values of sampling probabilities $p_{\alpha^{\text{s}}_{0}}$ and $p_{\alpha^{\text{s}}_{1}}$ are given by
		\begin{align}
			\label{optimal_sampling_prob1}
			p^{*}_{\alpha^{\text{s}}_{0}} = 0 , \hspace{0.2cm} p^{*}_{\alpha^{\text{s}}_{1}} = \text{min}\left\{1,\frac{\eta(p+q)}{p}\right\}.
		\end{align}
		This is because using \eqref{cond1}, as $p_{\alpha^{\text{s}}_{1}}< \frac{pC_{1,0}-qC_{0,1}}{p_{\text{s}_{1}}\big(pC_{1,0}+(1-q)C_{0,1}\big)}$, the objective function in \eqref{Optimization_prob2_objfun} is increasing with $p_{\alpha^{\text{s}}_{0}}$. Therefore, the optimal values of the sampling probabilities are given by \eqref{optimal_sampling_prob1}.
		\begin{remark}
			\label{remark_optimization}
			When $pC_{1,0}\geqslant qC_{0,1}$ and $p_{{\text{s}}_{1}}<\frac{pC_{1,0}-qC_{0,1}}{pC_{1,0}+(1-q)C_{0,1}}$, we cannot find a probability $p_{\alpha^{\text{s}}_{1}}\in [0,1]$ that satisfies the condition given in \eqref{cond1}. Therefore, in that case, the optimal values of $p_{\alpha^{\text{s}}_{0}}$ and $p_{\alpha^{\text{s}}_{1}}$ that minimize the objective function in \eqref{Optimization_prob2_objfun} are given by $p^{*}_{\alpha^{\text{s}}_{0}} = 0$ and $p^{*}_{\alpha^{\text{s}}_{1}} = \text{min}\left\{1,\frac{\eta(p+q)}{p}\right\}$.
		\end{remark}
		\item \emph{When $pC_{1,0}< qC_{0,1}$}: in this case, since $p_{\alpha^{\text{s}}_{1}}\geqslant 0$, using \eqref{cond1}, as $p_{\alpha^{\text{s}}_{0}}$ increases, the objective function in \eqref{Optimization_prob2_objfun} decreases. Using the constraint in \eqref{Optimization_prob2_constraint}, the maximum value of $p_{\alpha^{\text{s}}_{0}}$ is given by
		\begin{align}
			\label{pa0_optimal0}
			p_{\alpha^{\text{s}}_{0}}=\frac{\eta(p+q)-pp_{\alpha^{\text{s}}_{1}}}{q}.
		\end{align}
		Now, using \eqref{pa0_optimal0}, the optimization problem given in \eqref{Optimization_problem2} is simplified as
		\begin{subequations}
			\label{Optimization_problem4}
			\begin{align}
				&\underset{p_{\alpha^{\text{s}}_{0}}}{\text{minimize}}\hspace{0.3cm}\hspace{0.3cm} \frac{H}{K}\label{Optimization_problem4_objfunc}\\
				&\text{subject to}\hspace{0.3cm}	p^{\text{LB}}_{\alpha^{\text{s}}_{1}}\leqslant p_{\alpha^{\text{s}}_{1}}\leqslant 	p^{\text{UB}}_{\alpha^{\text{s}}_{1}} ,\label{Optimization_prob4_constraint}
			\end{align}
		\end{subequations}
		where $H$, $K$, $p^{\text{LB}}_{\alpha^{\text{s}}_{1}}$, and $p^{\text{UB}}_{\alpha^{\text{s}}_{1}}$ are given by
		\begin{align}
			\label{phi_2}
			H&\!=\!pq\bigg[C_{1,0}p_{{\text{s}}_{0}}\big(1-p_{\alpha^{\text{s}}_{1}}p_{{\text{s}}_{1}}\big)\big(pp_{\alpha^{\text{s}}_{1}}-\eta(p+q)\big)\notag\\
			&-C_{0,1}p_{\alpha^{\text{s}}_{1}}p_{{\text{s}}_{1}}\big(q+pp_{\alpha^{\text{s}}_{1}}p_{{\text{s}}_{0}}-\eta p_{{\text{s}}_{0}}(p+q)\big)\bigg]\notag\\
			K&\!=\!\big(p\!+\!q\big)\!\bigg[\eta p_{{\text{s}}_{0}}(p+q)\big(p_{\alpha^{\text{s}}_{1}}p_{{\text{s}}_{1}}(p\!+\!q\!-\!1)\!-\!q\big)\!\notag\\
			&-\!p p_{\alpha^{\text{s}}_{1}}\big(q(p_{{\text{s}}_{1}}\!-\!p_{{\text{s}}_{0}})\!+\!p_{{\text{s}}_{0}}p_{\alpha^{\text{s}}_{1}}p_{{\text{s}}_{1}}(p\!+\!q\!-\!1)\big)\bigg]\\
			p^{\text{LB}}_{\alpha^{\text{s}}_{1}} &= \text{max}\left\{\!0,\frac{\eta(p\!+\!q)\!-\!q}{p}\!\right\}\!,
			p^{\text{UB}}_{\alpha^{\text{s}}_{1}} \!=\!\text{min}\left\{\!1,\frac{\eta(p+q)}{p}\!\right\}\!.
		\end{align}
		Similar to the case when $pC_{1,0}\geqslant qC_{0,1}$, we can obtain $p_{\alpha^{\text{s}}_{1}}$ that minimizes the objective function in \eqref{Optimization_problem4_objfunc} by calculating its critical points within the interval $\big[	p^{\text{LB}}_{\alpha^{\text{s}}_{1}},	p^{\text{UB}}_{\alpha^{\text{s}}_{1}}\big]$. Then, we obtain $p_{\alpha^{\text{s}}_{0}}$ using \eqref{pa0_optimal0}. We can similarly prove that when $p_{\alpha^{\text{s}}_{0}}\geqslant \frac{qC_{0,1}-pC_{1,0}}{p_{\text{s}_{0}}\big(qC_{0,1}+(1-p)C_{1,0}\big)}$, $p_{\alpha^{\text{s}}_{0}}$ and $p_{\alpha^{\text{s}}_{1}}$ derived by solving the optimization problem in \eqref{Optimization_problem4}, are the optimal values of the sampling probabilities. Otherwise, the optimal values of the sampling probabilities are equal to $p^{*}_{\alpha^{\text{s}}_{0}} = \text{min}\left\{1,\frac{\eta(p+q)}{q}\right\}$ and $p^{*}_{\alpha^{\text{s}}_{1}} = 0$.
\end{enumerate}
\begin{remark}
	\label{remark_RSCdef}
	In what follows, RS and RSC policies are the abbreviations for the state-aware randomized stationary policy and the state-aware randomized stationary policy in the constrained optimization problem, respectively.
\end{remark}

\section{Simulation Results}
\label{simulation_results}
\par In this section, we validate our analytical results and evaluate the performance of the sampling policies in terms of time-averaged reconstruction error and the average cost of actuation error under various system parameters. In the uniform policy, a sample is acquired every 5
time slots. Simulation results are obtained averaging over $10^{7}$ time slots. 
\par In Tables \ref{Table1} through \ref{Table2_RS}, we illustrate the minimum average cost of actuation error when $C_{0,1} = 1$, $C_{1,0} = 2$ under a sampling cost constraint for $\eta = 0.5$, and various values of $p$, $q$, $p_{\text{s}_{0}}$, and $p_{\text{s}_{1}}$. As seen in these Tables, when $pC_{1,0}\geqslant qC_{0,1}$, the average cost of actuation error has its minimum value when $p_{\alpha^{\text{s}}_{1}}$ is greater than $p_{\alpha^{\text{s}}_{0}}$. Otherwise, the minimum average cost of actuation error occurs for $p_{\alpha^{\text{s}}_{0}}>p_{\alpha^{\text{s}}_{1}}$. Note also that for the lower successful probabilities, the minimum average cost of actuation error occurs at small values of $p^{*}_{\alpha^{\text{s}}_{0}}$ and $p^{*}_{\alpha^{\text{s}}_{1}}$, where $pC_{1,0}\geqslant q C_{0,1}$ and $qC_{0,1}> p C_{1,0}$, respectively. Furthermore, we observe that the optimal RSC policy exhibits superior performance than the semantics-aware policy under the conditions given in Remark \ref{remark_RS_SA_compare}, for both slow and fast changing the information source. Otherwise, when Remark \ref{remark_RS_SA_compare} is not satisfied, the semantics-aware policy performs better only when the source is slowly evolving. Note that the optimal values in red color for the semantics-aware, change-aware, and RS policies are obtained for values of $p$, $q$, $p_{\text{s}_{0}}$, and $p_{\text{s}_{1}}$ that do not satisfy the constraint requirement. This means that in the unconstrained scenario, the performance of the optimal RS policy is either better or the same as that of the semantics-aware policy. However, in this case, the optimal solution for the RS policy is to sample and transmit at most of the time slots, resulting in an excessive amount of samples being generated. 

The optimal values for sampling probabilities in the unconstrained scenario are shown in Tables \ref{Table1_RS} and \ref{Table2_RS}. As observed in these Tables, for all values of $p$ and $q$, we have $pC_{1,0}\geqslant q C_{0,1}$. Consequently, the optimal value of $p^{*}_{\alpha^{\text{s}}_{1}}$ is $1$. Furthermore, for values of $p$ and $q$ where $p_{\text{s}_{1}}<\frac{pC_{1,0}-qC_{0,1}}{pC_{1,0}+(1-q)C_{0,1}}$, the optimal value of $p^{*}_{\alpha^{\text{s}}_{0}}$ is $0$; otherwise, $p^{*}_{\alpha^{\text{s}}_{0}}=1$. This implies that when the success probability of a state is low, the optimal solution is to refrain from sampling for the state that causes the less important error in terms of actuation, while for a higher success probability, the optimal solution is to always perform sampling.
\begin{table}[]
	\centering
	\scriptsize
	\caption{\scriptsize{Minimum average cost of actuation error for RSC state-aware with $\eta = 0.5$, $C_{0,1} = 1$, $C_{1,0} = 2$, $p_{\text{s}_{0}} = 0.2$, $p_{\text{s}_{1}} = 0.3$, and different values of $p$ and $q$.}}
	\label{Table1}
	\begin{tabular}{|c|c|c|c|c|}
		\hline
		${p}$ &$q$& $p^{*}_{\alpha^{\text{s}}_{0}}$ & $p^{*}_{\alpha^{\text{s}}_{1}}$& $\text{{Minimum average cost of actuation error}}$\\
		\hline
		{0.1}  & {0.01}& {0.083}  & {0.542} & {0.091}\\ \hline
		{0.3} & {0.1}&{0}  & {0.667}  & {0.25}\\ \hline
		{0.5}  & {0.4}&{0}  & {0.9} & {0.444}\\ \hline
		{0.7} &{0.8}& {0}  & {1}  & {0.533}\\\hline
		{0.9} &{0.95}& {0}  & {1}  & {0.513}\\\hline
	\end{tabular}
\end{table}
\begin{table}[]
	\centering
	\scriptsize
	\caption{\scriptsize{Minimum average cost of actuation error for $\eta = 0.5$, $C_{0,1} = 1$, $C_{1,0} = 2$, $p_{\text{s}_{0}} = 0.2$, $p_{\text{s}_{1}} = 0.3$, and different values of $p$ and $q$.}}
	\label{Table_Optimal_M1}
	\begin{tabular}{|c|c|c|c|c|c|c|}
		\hline
		$p$ &$q$& $\text{Semantics-aware}$& $\text{Change-aware}$&$\text{Uniform}$&$\text{RSC}$&$\textcolor{red}{\text{RS}}$\\
		\hline
		0.1  &0.01& 0.055 & 0.628&0.131&0.091&\textcolor{red}{0.055}\\ \hline
		0.3 &0.1& 0.267  & 0.613 &0.417&0.25 &\textcolor{red}{0.25}\\ \hline
		0.5   &0.4& 0.489 & 0.596&0.638&0.444&\textcolor{red}{0.444}\\ \hline
		0.7  &0.8& \textcolor{red}{0.571}  & \textcolor{red}{0.588}&0.683&0.533&\textcolor{red}{0.533}\\ \hline
		0.9  &0.95& \textcolor{red}{0.587}  & \textcolor{red}{0.589}&0.677&0.513&\textcolor{red}{0.513} \\ \hline
	\end{tabular}
\end{table}

\begin{table}[]
	\centering
	\scriptsize
	\caption{\scriptsize{Minimum average cost of actuation error for RS state-aware with $\eta = 0.5$, $C_{0,1} = 1$, $C_{1,0} = 2$, $p_{\text{s}_{0}} = 0.2$, $p_{\text{s}_{1}} = 0.3$, and different values of $p$ and $q$.}}
	\label{Table1_RS}
	\begin{tabular}{|c|c|c|c|c|}
		\hline
		${p}$ &$q$& $p^{*}_{\alpha^{\text{s}}_{0}}$ & $p^{*}_{\alpha^{\text{s}}_{1}}$& $\text{{Minimum average cost of actuation error}}$\\
		\hline
		{0.1}  & {0.01}& {1}  & {1} & {0.055}\\ \hline
		{0.3} & {0.1}&{0}  & {1}  & {0.25}\\ \hline
		{0.5}  & {0.4}&{0}  & {1} & {0.444}\\ \hline
		{0.7} &{0.8}& {0}  & {1}  & {0.533}\\\hline
		{0.9} &{0.95}& {0}  & {1}  & {0.513}\\\hline
	\end{tabular}
\end{table}
\begin{table}[]
	\centering
	\scriptsize
	\caption{\scriptsize{Minimum average cost of actuation error for RSC state-aware with $\eta = 0.5$, $C_{0,1} = 1$, $C_{1,0} = 2$, $p_{\text{s}_{0}} = 0.6$, $p_{\text{s}_{1}} = 0.6$, and different values of $p$ and $q$.}}
	\label{Table2}
	\begin{tabular}{|c|c|c|c|c|}
		\hline
		${p}$ &$q$& $p^{*}_{\alpha^{\text{s}}_{0}}$ & $p^{*}_{\alpha^{\text{s}}_{1}}$& $\text{{Minimum average cost of actuation error}}$\\
		\hline
		{0.1}  & {0.01}& {0.730}  & {0.477} & {0.049}\\ \hline
		{0.3} & {0.1}&{0.155}  & {0.615}  & {0.241}\\ \hline
		{0.5}  & {0.4}&{0.171}  & {0.763} & {0.422}\\ \hline
		{0.7} &{0.8}& {0.200}  & {0.842}  & {0.501}\\\hline
		{0.9} &{0.95}& {0.127}  & {0.893}  & {0.503}\\\hline
	\end{tabular}
\end{table}
\begin{table}[]
	\centering
	\scriptsize
	\caption{\scriptsize{Minimum average cost of actuation error for $\eta = 0.5$, $C_{0,1} = 1$, $C_{1,0} = 2$, $p_{\text{s}_{0}} = 0.6$, $p_{\text{s}_{1}} = 0.6$, and different values of $p$ and $q$.}}
	\label{Table_Optimal_M2}
	\begin{tabular}{|c|c|c|c|c|c|c|}
		\hline
		$p$ &$q$& $\text{Semantics-aware}$& $\text{Change-aware}$&$\text{Uniform}$&$\text{RSC}$&$\textcolor{red}{\text{RS}}$\\
		\hline
		0.1  &0.01& 0.017 & 0.545&0.092&0.049&\textcolor{red}{0.017}\\ \hline
		0.3 &0.1& 0.118  & 0.5 &0.404&0.241 &\textcolor{red}{0.118}\\ \hline
		0.5   &0.4& 0.278 & 0.444&0.640&0.422&\textcolor{red}{0.278}\\ \hline
		0.7  &0.8& \textcolor{red}{0.373}  & \textcolor{red}{0.419}&0.686&0.501&\textcolor{red}{0.373}\\ \hline
		0.9  &0.95& \textcolor{red}{0.414}  & \textcolor{red}{0.424}&0.690&0.503&\textcolor{red}{0.414} \\ \hline
	\end{tabular}
\end{table}
\begin{table}[]
	\centering
	\scriptsize
	\caption{\scriptsize{Minimum average cost of actuation error for RS state-aware with $\eta = 0.5$, $C_{0,1} = 1$, $C_{1,0} = 2$, $p_{\text{s}_{0}} = 0.6$, $p_{\text{s}_{1}} = 0.6$, and different values of $p$ and $q$.}}
	\label{Table2_RS}
	\begin{tabular}{|c|c|c|c|c|}
		\hline
		${p}$ &$q$& $p^{*}_{\alpha^{\text{s}}_{0}}$ & $p^{*}_{\alpha^{\text{s}}_{1}}$& $\text{{Minimum average cost of actuation error}}$\\
		\hline
		{0.1}  & {0.01}& {1}  & {1} & {0.017}\\ \hline
		{0.3} & {0.1}&{1}  & {1}  & {0.118}\\ \hline
		{0.5}  & {0.4}&{1}  & {1} & {0.278}\\ \hline
		{0.7} &{0.8}& {1}  & {1}  & {0.373}\\\hline
		{0.9} &{0.95}& {1}  & {1}  & {0.414}\\\hline
	\end{tabular}
\end{table}
\par The performance of the optimal state-aware randomized stationary policy in terms of time-averaged reconstruction error as a function of $\eta$ for $p_{\text{s}_{0}} = 0.5$, $p_{\text{s}_{1}} = 0.6$, and different values of $p$, and $q$ is shown in Tables \ref{Table3} - \ref{Table_Optimal_M4}. We observe that the time-averaged reconstruction error has a smaller value as $\eta$ increases. This is because $\eta$ is the threshold of the total time-averaged sampling cost, thus a higher value of $\eta$ results in higher sampling probabilities which in turn decreases the time-averaged reconstruction error, as demonstrated by Tables \ref{Table_Optimal_M3} and \ref{Table_Optimal_M4}, $q>p$ and $p_{\text{s}_{0}}>\frac{q-p}{1+q-p}$. Hence, the minimum time-averaged reconstruction error in the unconstrained scenario is obtained with $p^{*}_{\alpha^{\text{s}}_{0}}=1$ and $p^{*}_{\alpha^{\text{s}}_{1}}=1$.
\begin{table}[]
	\centering
	\scriptsize
	\caption{\scriptsize{Minimum time-averaged reconstruction error as a function of $\eta$ for RSC state-aware with $p_{\text{s}_{0}} = 0.5$, $p_{\text{s}_{1}} = 0.6$, $p=0.2$ and $q=0.4$.}}
	\label{Table3}
	\begin{tabular}{|c|c|c|c|}
		\hline
		${\eta}$ & $p^{*}_{\alpha^{\text{s}}_{0}}$ & $p^{*}_{\alpha^{\text{s}}_{1}}$& $\text{{Minimum time-averaged reconstruction error}}$\\
		\hline
		{0.1}  & {0.15}& {0}  & {0.333}\\ \hline
		{0.3} & {0.394}&{0.112}  & {0.325}\\ \hline
		{0.5}  & {0.556}&{0.387}  & {0.277}\\ \hline
		{0.7} &{0.722}& {0.655}  & {0.224}\\\hline
		{0.9} &{0.889}& {0.922}  & {0.174}\\\hline
	\end{tabular}
\end{table}
\begin{table}[]
	\centering
	\scriptsize
	\caption{\scriptsize{Minimum time-averaged reconstruction error as a function of $\eta$ for $p_{\text{s}_{0}} = 0.5$, $p_{\text{s}_{1}} = 0.6$, $p=0.2$ and $q=0.4$.}}
	\label{Table_Optimal_M3}
	\begin{tabular}{|c|c|c|c|c|c|}
		\hline
		$\eta$ & $\text{Semantics-aware}$& $\text{Change-aware}$&$\text{Uniform}$&$\text{RSC}$&$\textcolor{red}{\text{RS}}$\\
		\hline
		0.1  & \textcolor{red}{0.151} & \textcolor{red}{0.333}&\textcolor{red}{0.374}&0.333&\textcolor{red}{0.151}\\ \hline
		0.3 & \textcolor{red}{0.151} & 0.333&0.374&0.325&\textcolor{red}{0.151}\\ \hline
		0.5   &{0.151} & {0.333}&{0.374}&0.277&\textcolor{red}{0.151}\\ \hline
		0.7  & {0.151} & {0.333}&{0.374}&0.224&\textcolor{red}{0.151}\\ \hline
		0.9  & {0.151} & {0.333}&{0.374}&0.174&\textcolor{red}{0.151} \\ \hline
	\end{tabular}
\end{table}
\begin{table}[]
	\centering
	\scriptsize
	\caption{\scriptsize{Minimum time-averaged reconstruction error for RSC state-aware with as a function of $\eta$ for $p_{\text{s}_{0}} = 0.5$, $p_{\text{s}_{1}} = 0.6$, $p=0.6$ and $q=0.7$.}}
	\label{Table4}
	\begin{tabular}{|c|c|c|c|}
		\hline
		${\eta}$ & $p^{*}_{\alpha^{\text{s}}_{0}}$ & $p^{*}_{\alpha^{\text{s}}_{1}}$& $\text{{Minimum time-averaged reconstruction error}}$\\
		\hline
		{0.1}  & {0.184}& {0.002}  & {0.461}\\ \hline
		{0.3} & {0.374}&{0.214}  & {0.430}\\ \hline
		{0.5}  & {0.565}&{0.424}  & {0.386}\\ \hline
		{0.7} &{0.757}& {0.633}  & {0.338}\\\hline
		{0.9} &{0.949}& {0.842}  & {0.287}\\\hline
	\end{tabular}
\end{table}
\begin{table}[]
	\centering
	\scriptsize
	\caption{\scriptsize{Minimum time-averaged reconstruction error as a function of $\eta$ for $p_{\text{s}_{0}} = 0.5$, $p_{\text{s}_{1}} = 0.6$, $p=0.6$ and $q=0.7$.}}
	\label{Table_Optimal_M4}
	\begin{tabular}{|c|c|c|c|c|c|}
		\hline
		$\eta$ & $\text{Semantics-aware}$& $\text{Change-aware}$&$\text{Uniform}$&$\text{RSC}$&$\textcolor{red}{\text{RS}}$\\
		\hline
		0.1  & \textcolor{red}{0.260} & \textcolor{red}{0.317}&\textcolor{red}{0.459}&0.461&\textcolor{red}{0.260}\\ \hline
		0.3 & \textcolor{red}{0.260} & \textcolor{red}{0.317}&0.459&0.430&\textcolor{red}{0.260}\\ \hline
		0.5   &\textcolor{red}{0.260} & \textcolor{red}{0.317}&{0.459}&0.386&\textcolor{red}{0.260}\\ \hline
		0.7  & {0.260} & {0.317}&{0.459}&0.338&\textcolor{red}{0.260}\\ \hline
		0.9  & {0.260} & {0.317}&{0.459}&0.287&\textcolor{red}{0.260} \\ \hline
	\end{tabular}
\end{table}
\par
\begin{figure}[!t]
	\centering
	\subfigure[$p_{{\text{s}}_{0}} = 0.2$, $p_{{\text{s}}_{1}}=0.3$]{\includegraphics[width=0.9\linewidth, clip]{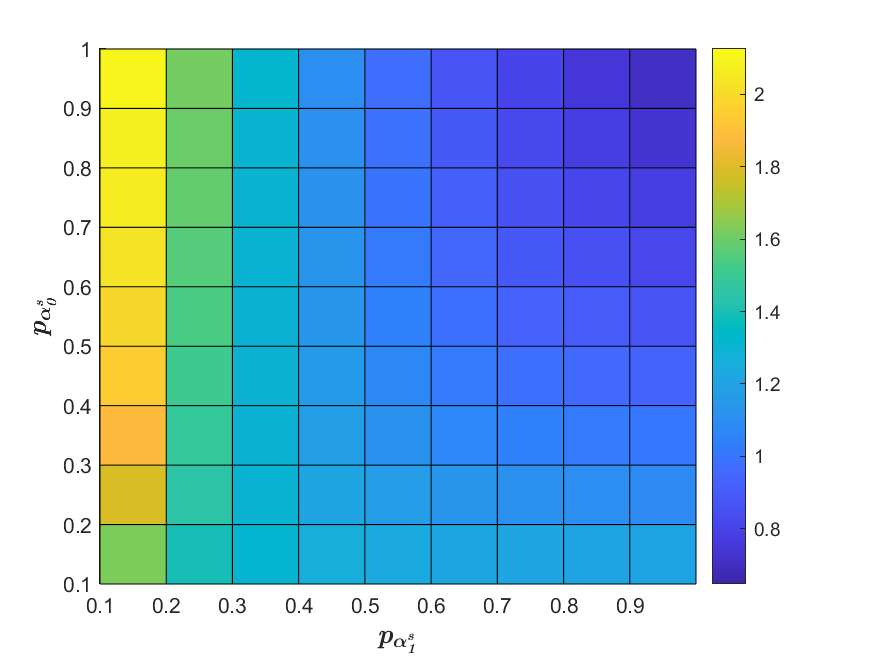}
	}
	\\
	\subfigure[$p_{{\text{s}}_{0}} = 0.7$, $p_{{\text{s}}_{1}}=0.8$]{\centering
		\includegraphics[width=0.9\linewidth, clip]{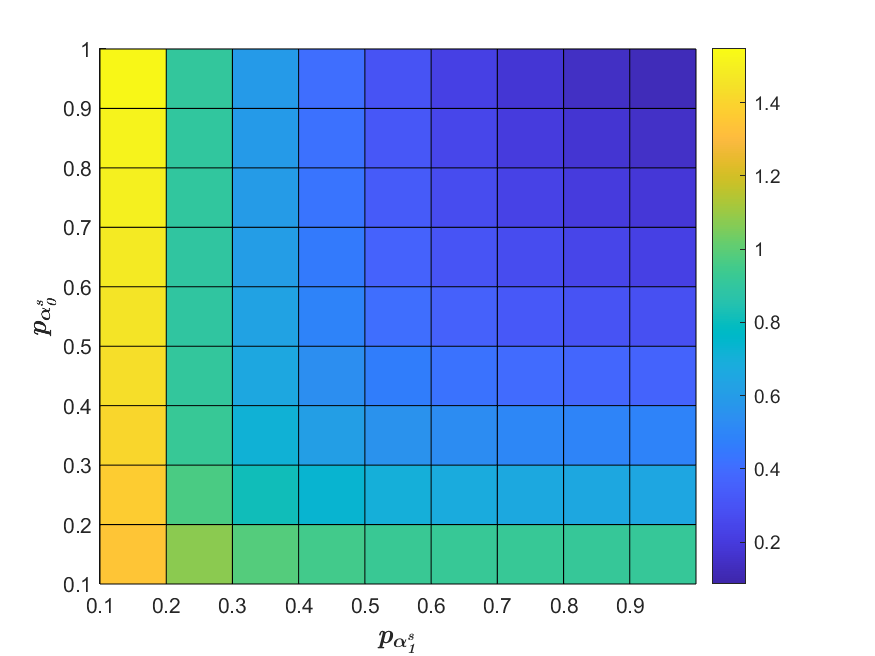}
	}
	\caption{Average consecutive error as a function of $p_{\alpha^{\text{s}}_{0}}$ and $p_{\alpha^{\text{s}}_{1}}$ for a slowly changing source with $p=0.3$, $q=0.2$.}
	\label{ConsecutiveError_slow}
\end{figure}

\begin{figure}[!t]
	\centering
	
	\subfigure[$p_{{\text{s}}_{0}} = 0.2$, $p_{{\text{s}}_{1}}=0.3$]{\includegraphics[width=0.9\linewidth, clip]{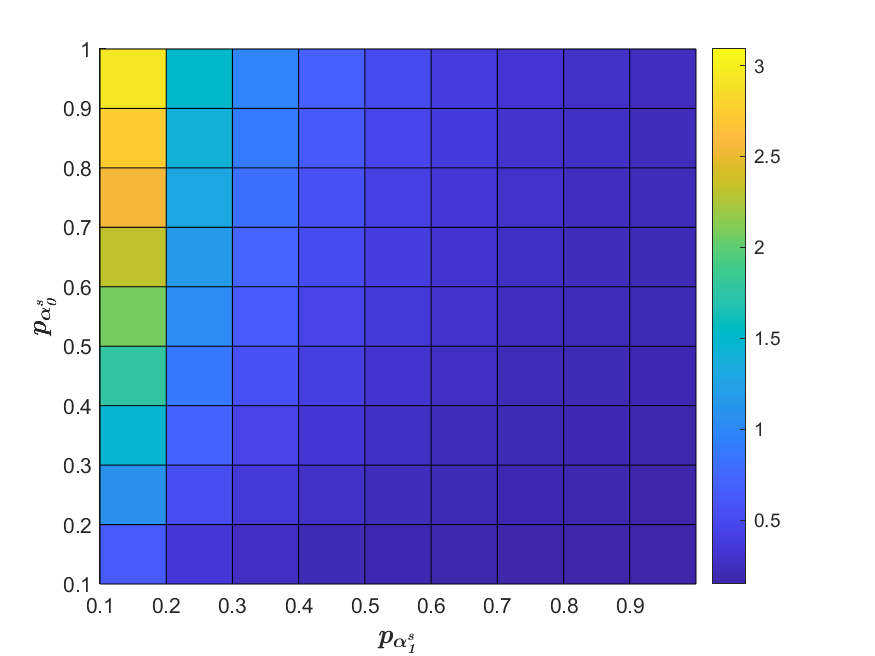}
	}
	\\
	\subfigure[$p_{{\text{s}}_{0}} = 0.7$, $p_{{\text{s}}_{1}}=0.8$]{\centering
		\includegraphics[width=0.9\linewidth, clip]{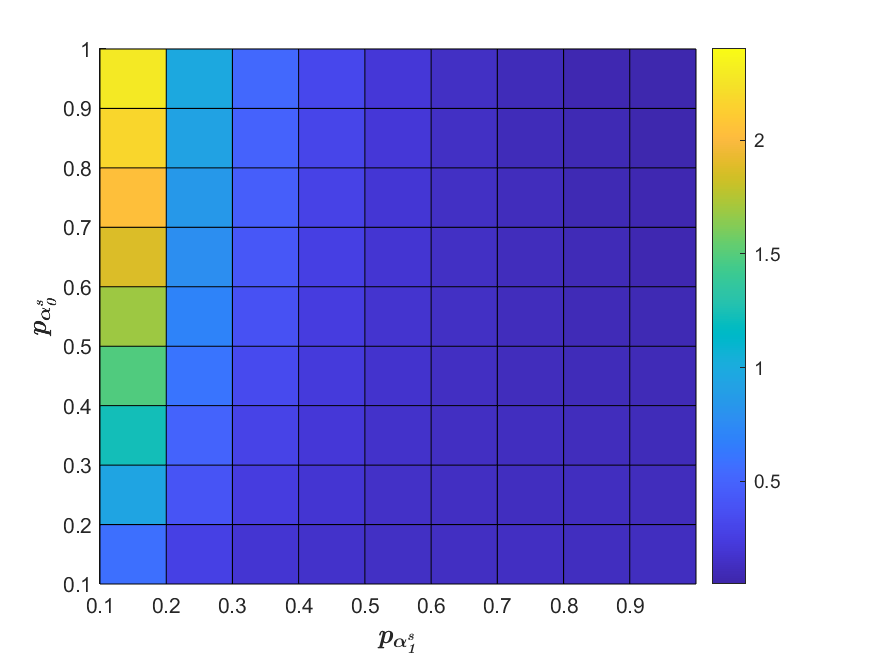}
	}
	\caption{Average consecutive error as a function of $p_{\alpha^{\text{s}}_{0}}$ and $p_{\alpha^{\text{s}}_{1}}$ for a rapidly changing source with $p=0.8$, $q=0.1$. }
	\label{ConsecutiveError_fast}
\end{figure}

 Figs. \ref{ConsecutiveError_slow}, and \ref{ConsecutiveError_fast} show the average consecutive error contour plots as a function of $p_{\alpha^{\text{s}}_{0}}$ and $p_{\alpha^{\text{s}}_{1}}$ for $p>q$, considering the slow and rapid changes of the source, respectively. As illustrated in Fig. \ref{ConsecutiveError_slow}, when the source changes slowly, the minimum average consecutive error occurs at high values of $p_{\alpha^{\text{s}}_{0}}$ and $p_{\alpha^{\text{s}}_{1}}$. In addition, as observed in Fig. \ref{ConsecutiveError_fast}, when the source changes rapidly and success probabilities are low, the average consecutive error decreases with a high value of $p_{\alpha^{\text{s}}_{1}}$ and a low value of $p_{\alpha^{\text{s}}_{0}}$. Furthermore, when success probabilities are high, the average consecutive error has its minimum value as $p_{\alpha^{\text{s}}_{0}}$ and $p_{\alpha^{\text{s}}_{1}}$ increase. Also, note that these figures can be used to obtain the optimal values of sampling probabilities. 
Another noteworthy result is that as the success probabilities increase, we can achieve a comparable average consecutive error with smaller sampling probabilities compared to situations where the success probabilities are lower. For example, when $p = 0.3$ and $q = 0.2$, with $p_{{\text{s}}_{0}} = 0.2$ and $p_{{\text{s}}_{1}} = 0.3$, the minimum average consecutive error is approximately $0.65$, which is achieved by setting $p_{\alpha^{\text{s}}_{0}} = 1$ and $p_{\alpha^{\text{s}}_{1}} = 1$. However, for $p_{{\text{s}}_{0}} = 0.7$ and $p_{{\text{s}}_{1}} = 0.8$, the similar average consecutive error value can be obtained by using $p_{\alpha^{\text{s}}_{0}} = 0.2$ and $p_{\alpha^{\text{s}}_{1}} = 1$.

\par
\begin{figure}[!t]
	\centering
	\subfigure[$p_{{\text{s}}_{0}} = 0.5$, $p_{{\text{s}}_{1}}=0.5$]{\includegraphics[width=0.9\linewidth, clip]{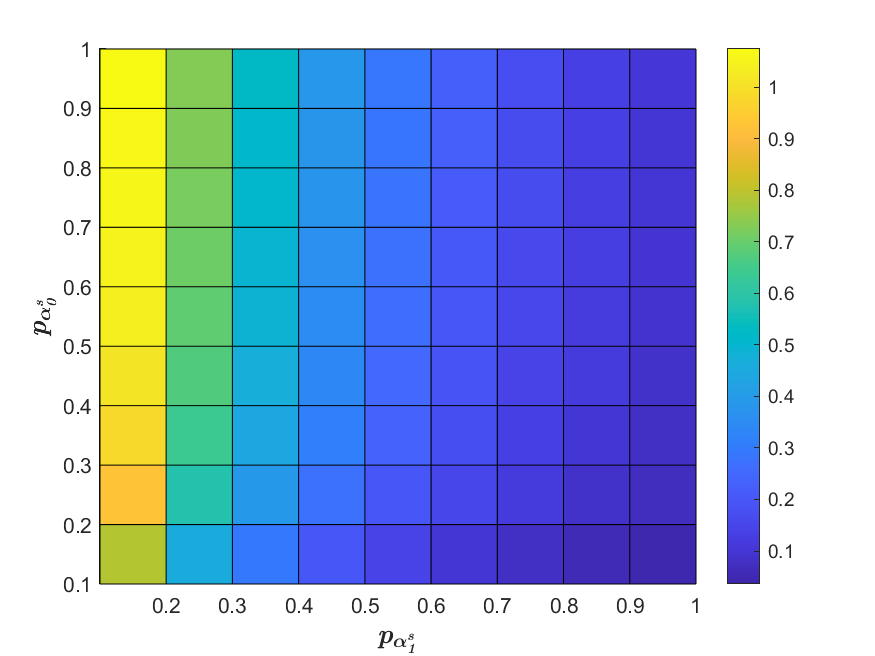}
	}
	\\
	\subfigure[$p_{{\text{s}}_{0}} = 0.9$, $p_{{\text{s}}_{1}}=0.8$]{\centering
		\includegraphics[width=0.9\linewidth, clip]{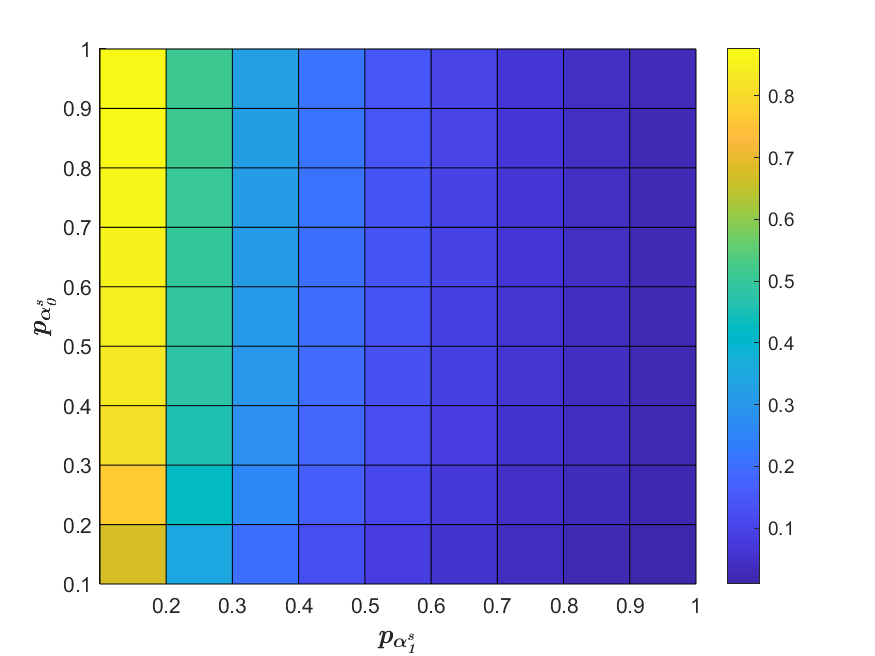}
	}
	\caption{Average importance-aware consecutive error as a function of $p_{\alpha^{\text{s}}_{0}}$ and $p_{\alpha^{\text{s}}_{1}}$ for a slowly changing source with $p=0.1$, $q=0.2$. }
	\label{Im_aware_ConsecutiveError_slow}
\end{figure}

\begin{figure}[!t]
	\centering
	
	\subfigure[$p_{{\text{s}}_{0}}=0.5$, $p_{{\text{s}}_{1}}=0.5$]{\includegraphics[width=0.9\linewidth, clip]{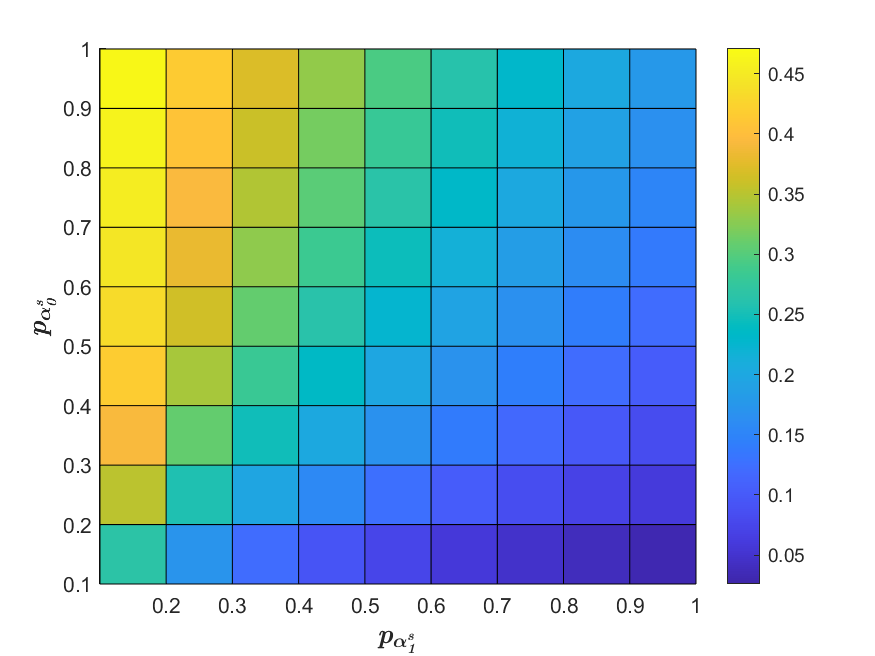}
	}
	\\
	\subfigure[$p_{{\text{s}}_{0}}=0.9$, $p_{{\text{s}}_{1}}=0.8$]{\centering
		\includegraphics[width=0.9\linewidth, clip]{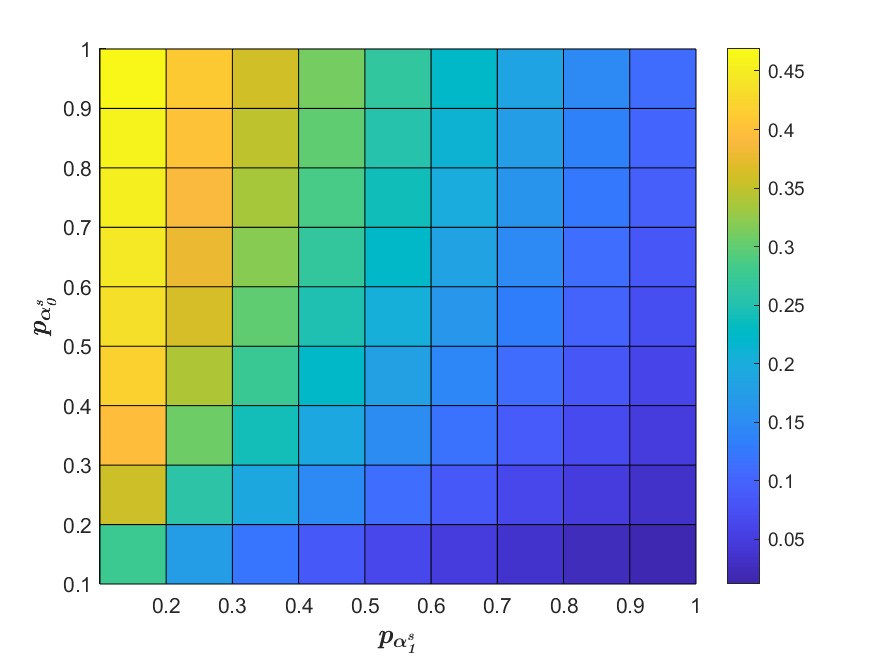}
	}
	\caption{Average importance-aware consecutive error as a function of $p_{\alpha^{\text{s}}_{0}}$ and $p_{\alpha^{\text{s}}_{1}}$ for a rapidly changing source with $p=0.8$, $q=0.9$.}
	\label{Im_aware_ConsecutiveError_fast}
\end{figure}

 The average importance-aware consecutive error as a function of $p_{\alpha^{\text{s}}_{0}}$ and $p_{\alpha^{\text{s}}_{1}}$ for slowly and rapidly changing information source, and selected values of success probabilities, is presented in Figs. \ref{Im_aware_ConsecutiveError_slow} and \ref{Im_aware_ConsecutiveError_fast}. In this analysis, we focus on the particular erroneous state where $X(t)=1$ and $\hat{X}(t)=0$. As seen in these figures, the average importance-aware consecutive error is minimized when $p_{\alpha^{\text{s}}_{1}}$ is at its maximum and $p_{\alpha^{\text{s}}_{0}}$ is at its minimum. The reason behind this is that when we have poor channel conditions, successful transmission of the less important state can have a negative impact since when the source will transit to the important state, the destination may miss that transition due to a potentially unsuccessful transmission. Thus, in that case, it may be preferable to avoid performing sampling in the less important state. To avoid such cases, this metric has to be considered and optimized in combination with other error metrics, such as the time-averaged reconstruction error. This is because a decrease in $p_{\alpha^{\text{s}}_{0}}$ may lead to an increase in the time-averaged reconstruction error, depending on the other system parameters. This increase can have significant consequences for the overall performance of such systems. Therefore, it is crucial to consider the interplay between the average importance-aware consecutive error and the time-averaged reconstruction error. Tables \ref{Table_Im_aware} and \ref{Table_PE} illustrate the average importance-aware consecutive error and time-averaged reconstruction error as a function of $p_{\alpha^{\text{s}}_{0}}$ and $p_{\alpha^{\text{s}}_{1}}$ for $p_{{\text{s}}_{0}} = 0.4$, $p_{{\text{s}}_{1}} = 0.7$, $p =0.5$, and $q=0.9$. Here we intend to find sampling probabilities $p_{\alpha^{\text{s}}_{0}}$ and $p_{\alpha^{\text{s}}_{1}}$ to keep the average importance-aware consecutive error and time-averaged reconstruction error below predefined  thresholds $I_{1}$ and $I_{2}$, respectively. For example, we assume that the thresholds are equal to $I_{1}=0.1$, and $I_{2} = 0.3$. As seen in these tables, we can achieve our purpose by setting $p_{\alpha^{\text{s}}_{0}}=1$, and $p_{\alpha^{\text{s}}_{1}}=0.9$ or $1$. However, when $I_{1}=0.1$, and $I_{2}=0.2$, we cannot find sampling probabilities $p_{\alpha^{\text{s}}_{0}}$ and $p_{\alpha^{\text{s}}_{1}}$  to fulfill our objective. While if we only consider the average importance-aware consecutive error metric, we could achieve our purpose by setting $p_{\alpha^{\text{s}}_{0}}=0.1$ and $p_{\alpha^{\text{s}}_{1}}=1$.
\begin{table}[!h]
	\centering
	\scriptsize
	\caption{\scriptsize{Average importance-aware consecutive error for $p_{\text{s}_{0}} = 0.4$, $p_{\text{s}_{1}} = 0.7$, $p=0.5$ and $q=0.9$, and selected values of $p_{\alpha^{\text{s}}_{0}}$ and $p_{\alpha^{\text{s}}_{1}}$.}}
	\label{Table_Im_aware}
	\begin{tabular}{|c|c|c|c|c|c|c|}
		\hline
		\diagbox{$p_{\alpha^{\text{s}}_{0}}$}{$p_{\alpha^{\text{s}}_{1}}$}&0.1 & 0.3& 0.5&0.7&0.9&1\\
		\hline
		0.1  & 0.189 &0.079&0.043&0.025&0.014&0.010\\ \hline
		0.3 &0.283 &0.163&0.101&0.063&0.037&0.028\\ \hline
		0.5  &0.315 &0.205&0.136&0.089&0.055&0.042\\ \hline
		0.7 &0.330  &0.231&0.161 &0.109&0.069&0.053\\ \hline
		0.9  &0.340&0.248 &0.179&0.124&0.081 &0.062\\ \hline
		1  & 0.343 &0.256 &0.186&0.131&0.086 &0.066\\ \hline
	\end{tabular}
\end{table}

\begin{table}[!h]
	\centering
	\scriptsize
	\caption{\scriptsize{Time-averaged reconstruction error for $p_{\text{s}_{0}} = 0.4$, $p_{\text{s}_{1}} = 0.7$, $p=0.5$ and $q=0.9$, and selected values of $p_{\alpha^{\text{s}}_{0}}$ and $p_{\alpha^{\text{s}}_{1}}$.}}
	\label{Table_PE}
	\begin{tabular}{|c|c|c|c|c|c|c|}
		\hline
		\diagbox{$p_{\alpha^{\text{s}}_{0}}$}{$p_{\alpha^{\text{s}}_{1}}$}&0.1 & 0.3& 0.5&0.7&0.9&1\\
		\hline
		0.1  & 0.480 &0.545&0.566&0.577&0.584&0.586\\ \hline
		0.3 &0.398 &0.443&0.466&0.480&0.490&0.494\\ \hline
		0.5  &0.371 &0.391&0.403&0.411&0.418&0.420\\ \hline
		0.7 &0.358  &0.358&0.359 &0.360&0.361&0.362\\ \hline
		0.9  &0.349&0.337 &0.328&0.321&0.314 &0.312\\ \hline
		1  & 0.346 &0.329 &0.315&0.304&0.294 &0.291\\ \hline
	\end{tabular}
\end{table}

\section{Conclusions}
\label{Conclusions}
\par In this work, we considered a time slotted communication system where sampling and transmission over a wireless erasure channel is performed in order to track a two-state Markov process. We proposed a state-aware randomized stationary, which considers varying sampling and success probabilities for different states of the source. We then analyzed the system performance in terms of set of metrics, namely time-averaged reconstruction error, average cost of actuation error, consecutive error, and importance-aware consecutive error. Furthermore, we cast and solved the optimization problem of minimizing the average cost of actuation error while keeping the time-averaged sampling cost constraint less than a given threshold. Our results illustrated that the optimal state-aware randomized stationary policy exhibits superior performance compared to other state-of-the-art policies when there is a constraint on the sampling cost, in particular when the source varies quickly. 
Specifically, it performs better for fast changes in the source and can also perform well for slow changes under certain conditions. 
\vspace{-.64 cm}
\begin{figure}[h!]
	\centering
	\scriptsize
	\begin{tikzpicture}[start chain=going left,->,>=latex,node distance=2.5cm]
		\node[state]    (A)                     {$(0,0)$};
		\node[state]    (B)[above right of=A]   {$(0,1)$};
		\node[state]    (C)[below right of=A]   {$(1,0)$};
		\node[state]    (D)[below right of=B]   {$(1,1)$};
		\path
		(A) edge[loop left]     node{$1-p$}  (A)
		edge[bend left=15,above]  node{$pp_{\alpha^{\text{s}}_{1}}p_{\text{s}_{1}}$}  (D)
		edge[bend right=15,left]    node{$P_{0}$}  (C)
		(B) edge[loop above]  node{$P_{1}$}     (B)
		edge[left=15] node{$(1-p)p_{\alpha^{\text{s}}_{0}}p_{\text{s}_{0}}$}    (A)
		edge[bend right=15,left] node{$p$}    (D)
		(C) edge[loop below]  node{$P_{2}$}     (C)
		edge[bend right=15,right] node{$q$}    (A)
		edge[ left=15,right] node{$(1-q)p_{\alpha^{\text{s}}_{1}}p_{\text{s}_{1}}$}    (D)
		(D) edge[loop right]    node{$1-q$}     (D)
		edge[bend right=15,right]  node{$P_{3}$}  (B)
		edge[bend left=15,above]     node{$qp_{\alpha^{\text{s}}_{0}}p_{\text{s}_{0}}$}         (A);
	\end{tikzpicture}
	\vspace*{1ex}
	\caption{Two-dimensional DTMC describing the joint status of the system regarding the current state at the original source using a two-state information source model.}
	\label{2DimMarkovChain}
\end{figure}
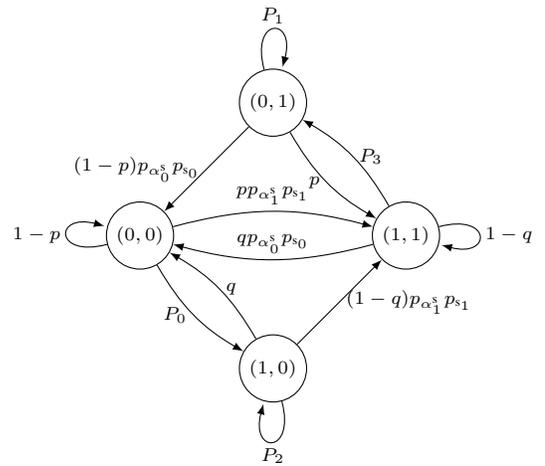
\appendix
\subsection{Proof of Lemma {\ref{lemma_PI_ij_DTMCmodel}}}
\label{Appendixlemma1}
\vspace{-.05in}
\par To obtain $\pi_{i,j}$ we depict the two-dimensional DTMC describing the joint status of the system regarding the current state at the original source, i.e., $(X(t), \hat{X}(t))$ in Fig. \ref{2DimMarkovChain}, where $P_{0}$, $P_{1}$, $P_{2}$, and $P_{3}$ are given by
\vspace{-0.07in}
\begin{align}
	\label{P0123}
	P_{0} &= \mathrm{Pr}\big[X_{t+1}=1, \hat{X}_{t+1}=0\big|X_{t}=0, \hat{X}_{t}=0\big] \notag\\
	&= pp_{\alpha^{\text{s}}_{1}}(1-p_{{\text{s}}_{1}})+p(1-p_{\alpha^{\text{s}}_{1}})\notag\\
	P_{1} &= \mathrm{Pr}\big[X_{t+1}=0, \hat{X}_{t+1}=1\big|X_{t}=0, \hat{X}_{t}=1\big]\notag\\
	&=(1-p)p_{\alpha^{\text{s}}_{0}}(1-p_{{\text{s}}_{0}})+(1-p)(1-p_{\alpha^{\text{s}}_{0}})\notag\\
	P_{2} &= \mathrm{Pr}\big[X_{t+1}=1, \hat{X}_{t+1}=0\big|X_{t}=1, \hat{X}_{t}=0\big]\notag\\
	&=(1-q)p_{\alpha^{\text{s}}_{1}}(1-p_{{\text{s}}_{1}})+(1-q)(1-p_{\alpha^{\text{s}}_{1}})\notag\\
	P_{3} &= \mathrm{Pr}\big[X_{t+1}=0, \hat{X}_{t+1}=1\big|X_{t}=1, \hat{X}_{t}=1\big]\notag\\
	&=
	qp_{\alpha^{\text{s}}_{0}}(1-p_{{\text{s}}_{0}})+q(1-p_{\alpha^{\text{s}}_{0}}).
\end{align}

Now, using Fig. \ref{2DimMarkovChain} and \eqref{P0123}, we can obtain state stationary $\pi_{i,j}, \forall i,j\in \{0,1\}$, as follows
\begin{align}
	\label{pij}
	\pi_{0,0}&=\frac{q p_{\alpha^{\text{s}}_{0}} p_{{\text{s}}_{0}} \big[q+(1-q)p_{\alpha^{\text{s}}_{1}} p_{{\text{s}}_{1}}\big]}{(p+ q)\Phi\big(p_{\alpha^{\text{s}}_{0}},p_{\alpha^{\text{s}}_{1}}\big)}\notag\\
	\pi_{0,1}&=\frac{pq p_{\alpha^{\text{s}}_{1}} p_{{\text{s}}_{1}} \big(1-p_{\alpha^{\text{s}}_{0}} p_{{\text{s}}_{0}}\big)}{(p+ q)\Phi\big(p_{\alpha^{\text{s}}_{0}},p_{\alpha^{\text{s}}_{1}}\big)}\notag\\
	\pi_{1,0}&=\frac{pq p_{\alpha^{\text{s}}_{0}} p_{{\text{s}}_{0}} \big(1-p_{\alpha^{\text{s}}_{1}} p_{{\text{s}}_{1}}\big)}{(p+ q)\Phi\big(p_{\alpha^{\text{s}}_{0}},p_{\alpha^{\text{s}}_{1}}\big)}\notag\\
	\pi_{1,1}&=\frac{p p_{\alpha^{\text{s}}_{1}} p_{{\text{s}}_{1}} \big[p+(1-p)p_{\alpha^{\text{s}}_{0}} p_{{\text{s}}_{0}}\big]}{(p+ q)\Phi\big(p_{\alpha^{\text{s}}_{0}},p_{\alpha^{\text{s}}_{1}}\big)},
\end{align}
where $\Phi\big(p_{\alpha^{\text{s}}_{0}},p_{\alpha^{\text{s}}_{1}}\big)$ is given in \eqref{Dnum}. For the change-aware policy, \eqref{pij} can be written as
\begin{align}
	\label{pij_changeaware}
	\pi_{0,0}&=\frac{q p_{{\text{s}}_{0}}}{(p+ q)\big(p_{{\text{s}}_{0}}+p_{{\text{s}}_{1}}-p_{{\text{s}}_{0}}p_{{\text{s}}_{1}}  \big)},\notag\\ \pi_{0,1}&=\frac{q p_{{\text{s}}_{1}}\big(1-p_{{\text{s}}_{0}}\big)}{(p+ q)\big( p_{{\text{s}}_{0}}+p_{{\text{s}}_{1}}-p_{{\text{s}}_{0}}p_{{\text{s}}_{1}}  \big)}\notag\\
	\pi_{1,0}&=\frac{p p_{{\text{s}}_{0}}\big(1-p_{{\text{s}}_{1}}\big)}{(p+ q)\big( p_{{\text{s}}_{0}}+p_{{\text{s}}_{1}}-p_{{\text{s}}_{0}}p_{{\text{s}}_{1}}  \big)}\notag\\
	\pi_{1,1}&=\frac{p p_{{\text{s}}_{1}}}{(p+ q)\big( p_{{\text{s}}_{0}}+p_{{\text{s}}_{1}}-p_{{\text{s}}_{0}}p_{{\text{s}}_{1}}  \big)}.
\end{align}
Furthermore, for the semantics-aware policy $\pi_{i,j}$ is given by
\begin{align}
	\label{pij_SA}
	\pi_{0,0}&=\frac{q p_{{\text{s}}_{0}} \big[q+(1-q) p_{{\text{s}}_{1}}\big]}{(p+q)\big[qp_{{\text{s}}_{0}}+(1-q)p_{{\text{s}}_{0}}p_{{\text{s}}_{1}}+pp_{{\text{s}}_{1}}(1-p_{{\text{s}}_{0}})\big]}\notag\\
	\pi_{0,1}&=\frac{pq  p_{{\text{s}}_{1}} \big(1- p_{{\text{s}}_{0}}\big)}{(p+q)\big[qp_{{\text{s}}_{0}}+(1-q)p_{{\text{s}}_{0}}p_{{\text{s}}_{1}}+pp_{{\text{s}}_{1}}(1-p_{{\text{s}}_{0}})\big]}\notag\\
	\pi_{1,0}&=\frac{pq p_{{\text{s}}_{0}} \big(1- p_{{\text{s}}_{1}}\big)}{(p+q)\big[qp_{{\text{s}}_{0}}+(1-q)p_{{\text{s}}_{0}}p_{{\text{s}}_{1}}+pp_{{\text{s}}_{1}}(1-p_{{\text{s}}_{0}})\big]}\notag\\
	\pi_{1,1}&=\frac{p p_{{\text{s}}_{1}} \big[p+(1-p) p_{{\text{s}}_{0}}\big]}{(p+q)\big[qp_{{\text{s}}_{0}}+(1-q)p_{{\text{s}}_{0}}p_{{\text{s}}_{1}}+pp_{{\text{s}}_{1}}(1-p_{{\text{s}}_{0}})\big]}.
\end{align}
%%%%%%%%%%%%%%%%%%%%%%%%%%%%%%%%%%%%%%%%%%%%%%
\subsection{Three-state DTMC information source}
\label{3States}
\par
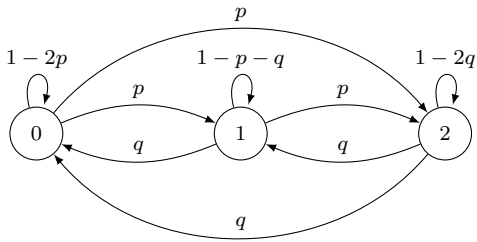
\begin{figure}[!h]
	\centering
	\footnotesize
	{  
		\begin{tikzpicture}[start chain=going left,->,>=latex,node distance=2cm]
			\footnotesize
			\node[state, on chain]        
			(3){$2$};
			
			\node[state, on chain]                 (2){$1$};
			\node[state, on chain]                 (1){$0$};
			\draw[>=latex]
			
			(1)   edge[loop above] node {$1-2p$} ()
			(1)   edge[bend left=23] node[above] {$p$}(2)
			(1)   edge[bend left=50] node[above]{$p$} (3)
			(2)   edge[loop above] node {$1-p-q$} ()
			(2)   edge[bend left=23] node[above] {$q$}(1)
			(2)   edge[bend left=23] node[above]{$p$} (3)
			(3)   edge[loop above] node {$1-2q$} ()
			(3)   edge[bend left=23] node[above] {$q$}(2)
			(3)   edge[bend left=50] node[above] {$q$}(1)
			;
		\end{tikzpicture}
	}
	\vspace*{1ex}
	\caption{The three-state DTMC model describing the evolution of the information source.}
	\label{3States_Infor_Source_DTMC}
\end{figure}

 For a three-state DTMC information source model depicted in Fig. \ref{3States_Infor_Source_DTMC}, the stationary distribution $\pi_{i,j}$ for the state-aware randomized stationary policy is given by
	\begin{align}
		\pi_{0,0}&=\frac{1}{Z_{1}}\Big[q^2(3p-1)p_{\alpha_{0}}p_{\text{s}_{0}}p_{\alpha_{1}}p_{\text{s}_{1}}\Big(p\big(p_{\alpha_{2}}p_{\text{s}_{2}}-1\big)-2q\notag\\
		&+(2q\!-\!1)p_{\alpha_{2}}p_{\text{s}_{2}}\Big)\Big] \Big[p\big(p_{\alpha_{1}}p_{\text{s}_{1}}\!-\!1\big)\big((q\!-\!1)p_{\alpha_{2}}p_{\text{s}_{2}}\!-\!q\big)\notag\\
		&+\big(p_{\alpha_{1}}p_{\text{s}_{1}}(q-1)-q\big)\big((2q-1)p_{\alpha_{2}}p_{\text{s}_{2}}-2q\big)\Big],\label{pi00_3}\\
		\pi_{0,1}&=\frac{1}{Z_{1}}\Big[pq^2(3p-1)p_{\alpha_{1}}p_{\text{s}_{1}}\big(p_{\alpha_{0}}p_{\text{s}_{0}}-1\big)\Big(p\big(p_{\alpha_{2}}p_{\text{s}_{2}}-1\big)\notag\\
		&-2q+(2q-1)p_{\alpha_{2}}p_{\text{s}_{2}}\Big)\Big]\Big[2pp_{\alpha_{2}}p_{\text{s}_{2}}\big(p_{\alpha_{1}}p_{\text{s}_{1}}-1\big)\notag\\
		&-qp_{\alpha_{2}}p_{\text{s}_{2}}+p_{\alpha_{1}}p_{\text{s}_{1}}\big(4qp_{\alpha_{2}}p_{\text{s}_{2}}-2p_{\alpha_{2}}p_{\text{s}_{2}}-3q\big)\Big],\label{pi01_3}\\
		\pi_{0,2}&=\frac{1}{Z_{2}}\Big[\!pqp_{\alpha_{2}}p_{\text{s}_{2}}\Big(p\big(\!p_{\alpha_{1}}p_{\text{s}_{1}}-1\!\big)\!-\!2q\!+\!(2q\!-\!1)p_{\alpha_{1}}p_{\text{s}_{1}}\Big)\!\Big],\label{pi02}\\
		\pi_{1,0}&=\frac{1}{Z_{1}}\Big[\!pq^2(3p\!-\!1)p_{\alpha_{0}}p_{\text{s}_{0}}p_{\alpha_{1}}p_{\text{s}_{1}}\!\big(p_{\alpha_{1}}p_{\text{s}_{1}}\!\!\!-\!\!1\big)\Big(\!p\big(p_{\alpha_{2}}p_{\text{s}_{2}}\!\!-\!\!1\big)\notag\\
		&+(2q-1)p_{\alpha_{2}}p_{\text{s}_{2}}-2q\Big)\Big((3q-1)p_{\alpha_{2}}p_{\text{s}_{2}}-3q\Big)\!\Big],\label{pi10_3}\\
		\pi_{1,1}&=\frac{1}{Z_{1}}\Big[pq(3p-1)p_{\alpha_{1}}p_{\text{s}_{1}}\Big(2pp_{\alpha_{2}}p_{\text{s}_{2}}\big(p_{\alpha_{1}}p_{\text{s}_{1}}-1\big)\notag\\
		&-qp_{\alpha_{2}}p_{\text{s}_{2}}+p_{\alpha_{1}}p_{\text{s}_{1}}\big(4qp_{\alpha_{2}}p_{\text{s}_{2}}-2p_{\alpha_{2}}p_{\text{s}_{2}}-3q\big)\Big)\Big]\notag\\
		&\times\Big[p\big(p_{\alpha_{0}}p_{\text{s}_{0}}-1\big)\big(-3q+(3q-2)p_{\alpha_{2}}p_{\text{s}_{2}}\big)\notag\\
		&+p_{\alpha_{0}}p_{\text{s}_{0}}\big(2q+p_{\alpha_{2}}p_{\text{s}_{2}}(1-2q)\big)\Big],\label{pi11_3}\\
		\pi_{1,2}&=\frac{1}{Z_{2}}\Big[pqp_{\alpha_{2}}p_{\text{s}_{2}}(1-3p)\big(1-p_{\alpha_{1}}p_{\text{s}_{1}}\big)\Big],\label{pi12_3}\\
		\pi_{2,0}&=\frac{1}{Z_{1}}\Big[pq^2(3p-1)p_{\alpha_{0}}p_{\text{s}_{0}}p_{\alpha_{1}}p_{\text{s}_{1}}\big(p_{\alpha_{2}}p_{\text{s}_{2}}-1\big)\notag\\
		&\times\Big(2p\big(p_{\alpha_{1}}p_{\text{s}_{1}}-1\big)+(q-1)p_{\alpha_{1}}p_{\text{s}_{1}}-q\Big)\notag\\
		&\times\Big(p\big(p_{\alpha_{2}}p_{\text{s}_{2}}-1\big)-2q+(2q-1)p_{\alpha_{2}}p_{\text{s}_{2}}\Big)\Big],\label{pi20_3}\\
		\pi_{2,1}&=\frac{1}{Z_{1}}\Big[qp^2(3p\!-\!1)p_{\alpha_{1}}p_{\text{s}_{1}}\big(p_{\alpha_{2}}p_{\text{s}_{2}}\!\!-\!\!1\big)\Big(2p\big(p_{\alpha_{0}}p_{\text{s}_{0}}-1\big)\notag\\&+(q\!-\!1)p_{\alpha_{0}}p_{\text{s}_{0}}\!-\!q\Big)\Big]\Big[2pp_{\alpha_{2}}p_{\text{s}_{2}}\big(p_{\alpha_{1}}p_{\text{s}_{1}}\!-\!1\big)\!-\!qp_{\alpha_{2}}p_{\text{s}_{2}}\notag\\
		&+p_{\alpha_{1}}p_{\text{s}_{1}}\big(4qp_{\alpha_{2}}p_{\text{s}_{2}}-2p_{\alpha_{2}}p_{\text{s}_{2}}-3q\big)\Big],\label{pi21_3}\\
		\pi_{2,2}&=\frac{1}{Z_{2}}\Big[pp_{\alpha_{2}}p_{\text{s}_{2}}\Big(p+2p^2\big(p_{\alpha_{1}}p_{\text{s}_{1}}-1\big)+p(q-3)p_{\alpha_{1}}p_{\text{s}_{1}}\notag\\
		&+q-pq+p_{\alpha_{1}}p_{\text{s}_{1}}(1-q)\Big)\Big]\label{pi22_3}.
	\end{align}
	where $Z_{1}$ and $Z_{2}$ in eqs. \eqref{pi00_3} to \eqref{pi22_3} are given by
	\begin{align}
		Z_{1} &= (2p+q)\Big[2p^3p_{{\alpha}_2}p_{\text{s}_2}\big(p_{{\alpha}_1}p_{\text{s}_{1}}-1\big)\notag\\
		&+qp_{\alpha_{1}}p_{\text{s}_{1}}\big(2q+p_{\alpha_2}p_{\text{s}_2}(1-2q)\big)+p^2\Big(-3qp_{\alpha_{1}}p_{\text{s}_{1}}\notag\\&+p_{\alpha_{2}}p_{\text{s}_{2}}\big(1\!-\!5q\!+\!p_{\alpha_{1}}p_{\text{s}_{1}}(8q\!-\!3)\big)\Big)\!+\!p\Big(\!-\!2q(q\!-\!1)p_{\alpha_{2}}p_{\text{s}_{2}}\notag\\
		&+p_{\alpha_{1}}p_{\text{s}_{1}}\big(q-6q^2+p_{\alpha_{2}}p_{\text{s}_{2}}\big(1-7q+8q^2\big)\big)\Big)\Big]\notag\\
		&\times\Big[2p^2p_{\alpha_{2}}p_{\text{s}_{2}}\big(p_{\alpha_{0}}p_{\text{s}_{0}}-1\big)\big(p_{\alpha_{1}}p_{\text{s}_{1}}-1\big)\notag\\
		&+p_{\alpha_{0}}p_{\text{s}_{0}}\big(p_{\alpha_{1}}p_{\text{s}_{1}}(q-1)-q\big)\big(-2q+p_{\alpha_{2}}p_{\text{s}_{2}}(2q-1)\big)\notag\\
		&+p\Big(qp_{\alpha_{2}}p_{\text{s}_{2}}+p_{\alpha_{1}}p_{\text{s}_{1}}\big(3q+(2-4q)p_{\alpha_{2}}p_{\text{s}_{2}}\big)\notag\\	&\!+\!p_{\alpha_{0}}p_{\text{s}_{0}}\!\big(q\!-\!4qp_{\alpha_{1}}p_{\text{s}_{1}}\!+\!p_{\alpha_{2}}p_{\text{s}_{2}}\big(1\!-\!2q\!+\!p_{\alpha_{1}}p_{\text{s}_{1}}(5q\!-\!3)\big)\big)\Big)\Big]\!,
	\end{align}
	\begin{align}
		Z_{2}&=2p^3p_{\alpha_{2}}p_{\text{s}_{2}}\big(p_{\alpha_{1}}p_{\text{s}_{1}}\!-\!1\big)\!+\!qp_{\alpha_{1}}p_{\text{s}_{1}}\big(2q\!+\!p_{\alpha_{2}}p_{\text{s}_{2}}(1\!-\!2q)\big)\notag\\&+p^2\Big(-3qp_{\alpha_{1}}p_{\text{s}_{1}}+p_{\alpha_{2}}p_{\text{s}_{2}}\big(1-5q+p_{\alpha_{1}}p_{\text{s}_{1}}(8q-3)\big)\Big)\notag\\
		&\!+\!p\Big(\!2q(1\!-\!q)p_{\alpha_{2}}p_{\text{s}_{2}}\!\!+\!\!p_{\alpha_{1}}p_{\text{s}_{1}}\big(q\!-\!6q^2\!\!+\!\!p_{\alpha_{2}}p_{\text{s}_{2}}(1\!-\!7q\!+\!8q^2)\big)\!\Big)\!.
	\end{align}

\subsection{Proof of Equation \eqref{Pr_CE_i}}
\label{Proof_P01_P11_RS}
\par $C_{E}(t)=i\hspace{0.05cm}  (i\geqslant 1)$ means that the system was in a synced state at time slot $t-i$, and it has been in an erroneous state from time slots $t-i+1$ to $t$. Therefore, to obtain $\mathrm{Pr}\big[C_{E}(t)=i\big]$, we need to calculate
\begin{align}
	\label{PCEi_proof}
	&\mathrm{Pr}\big[C_{E}(t)=i\big]\notag\\
	&=\mathrm{Pr}\big[E(t)\!\neq\! 0,\!\cdots\!,  E(t\!-\!i\!+\!1)\!\neq \!0, E(t\!-\!i)\!=\!0\big]\notag\\
	&=\mathrm{Pr}\big[E(t)\!\neq\! 0,\!\cdots\!,  E(t\!-\!i\!+\!1)\!\neq \!0\big|X(t\!-\!i)\!=\!0, E(t\!-\!i)\!=\!0\big]\notag\\
	&\times\mathrm{Pr}\big[X(t-i)=0, E(t-i)=0\big]\notag\\
	&+\mathrm{Pr}\big[E(t)\!\neq\! 0,\!\cdots\!,  E(t\!-\!i\!+\!1)\!\neq \!0\big|X(t\!-\!i)\!=\!1, E(t\!-\!i)\!=\!0\big]\notag\\
	&\times\mathrm{Pr}\big[X(t-i)=1, E(t-i)=0\big],
\end{align}
where the first conditional probability in \eqref{PCEi_proof} can be written as
\begin{align}
	\label{PCEi_proof_1}
	&\mathrm{Pr}\big[E(t)\!\neq\! 0,\!\cdots\!,  E(t\!-\!i\!+\!1)\!\neq \!0\big|X(t\!-\!i)\!=\!0, E(t\!-\!i)\!=\!0\big]\notag\\
	&=\!\mathrm{Pr}\big[X(t\!-\!i\!+\!1)\!=\!1,\hat{X}(t\!-\!i\!+\!1)\!=\!0\big|X(t\!-\!i)\!=\!0, \hat{X}(t\!-\!i)\!=\!0\big]\notag\\
	&\!\times\!\big(i\geqslant 2\big)\Bigg\{\prod_{j=1-i}^{-1}\mathrm{Pr}\big[X(t\!+\!j\!+\!1)\!=\!1,\hat{X}(t\!+\!j\!+\!1)\!=\!0\notag\\
	&\hspace{0.5cm}\big|X(t\!+\!j)\!=\!1, \hat{X}(t\!+\!j)\!=\!0\big]\Bigg\}
	=p(1-q)^{i-1}\big(1-p_{\alpha^{\text{s}}_{1}}p_{{\text{s}}_{1}}\big)^{i}.
\end{align}
Similarly, one can obtain the second conditional probability in \eqref{PCEi_proof} as
\begin{align}
	\label{PCEi_proof_2}
	&\mathrm{Pr}\big[E(t)\!\neq\! 0,\!\cdots\!,  E(t\!-\!i\!+\!1)\!\neq \!0\big|X(t\!-\!i)\!=\!1, E(t\!-\!i)\!=\!0\big]\notag\\
	&
	=q(1-p)^{i-1}\big(1-p_{\alpha^{\text{s}}_{0}}p_{{\text{s}}_{0}}\big)^{i}.
\end{align}
Now, using Lemma {\ref{lemma_PI_ij_DTMCmodel}}, \eqref{PCEi_proof_1} and \eqref{PCEi_proof_2}, we can write \eqref{PCEi_proof} as
\begin{align}
	\label{Pr_CE_i_Proof_final}
	&\mathrm{Pr}\big[C_{E}(t)=i\big] \notag\\
	&\!\!=\!p(1\!-\!q)^{i-1}\big(1\!-\!p_{\alpha^{\text{s}}_{1}} p_{{\text{s}}_{1}}\big)^{i}\!\pi_{0,0}\!+\!q(1\!-\!p)^{i-1}\big(1\!-\!p_{\alpha^{\text{s}}_{0}} p_{{\text{s}}_{0}}\big)^{i}\!\pi_{1,1}.
\end{align}

\bibliographystyle{IEEEtran}
\bibliography{ref}
\begin{IEEEbiography}[{\includegraphics[width=1.1in,height=1.25in,clip,keepaspectratio]{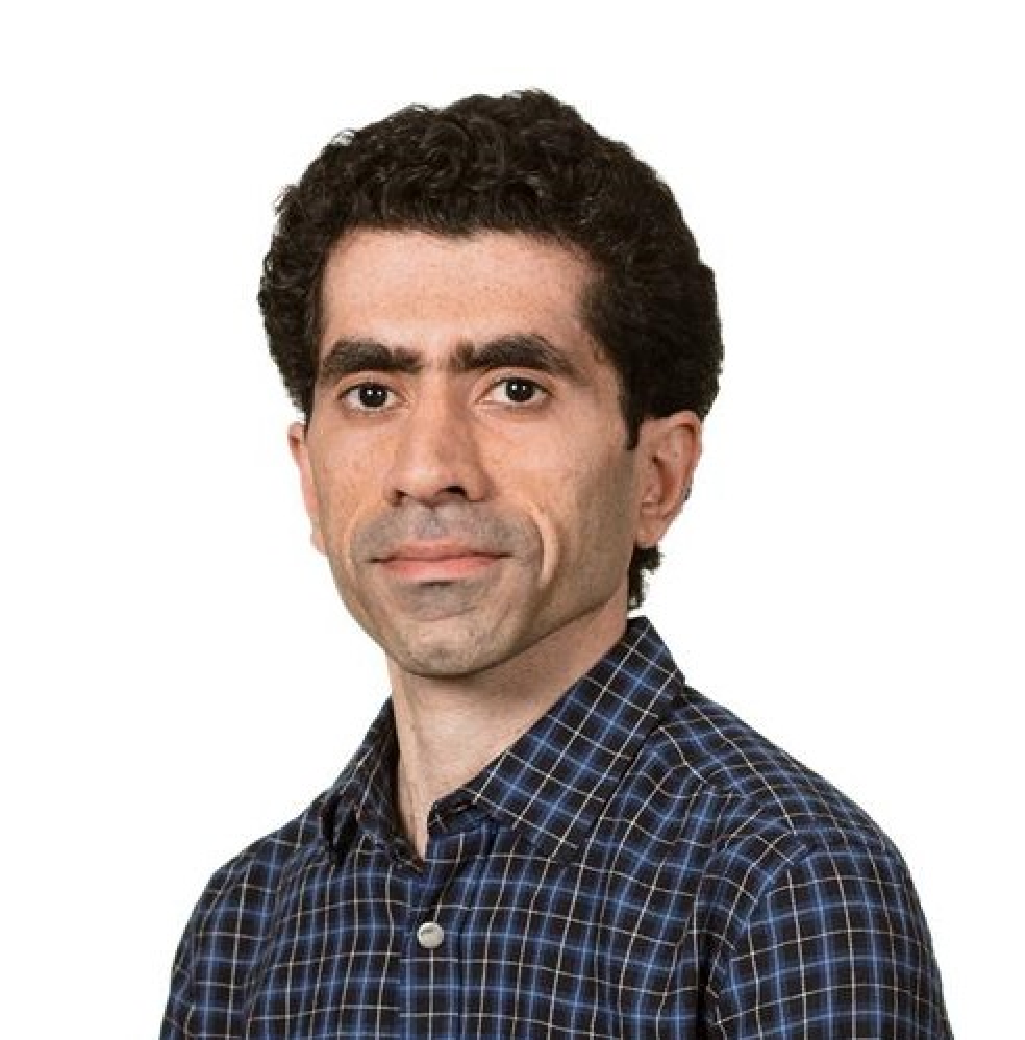}}]{Mehrdad Salimnejad}(Student Member, IEEE)
	received the B.Sc. degree in electrical engineering from the Razi University, Kermanshah, Iran, in 2012, and the M.Sc. degree in electrical engineering from the University of Tehran, Tehran, Iran, in 2015.
	\par From 2015 to 2022, he was a Research Engineer at the Research Center of Sharif University of Technology working on the design and development of the fifth-generation (5G) wireless cellular networks. He is currently Ph.D. student at the Department of Computer and Information Science at Linköping University, Sweden. His interest include semantic wireless communication, age of information, and communication networks.
\end{IEEEbiography}

\begin{IEEEbiography}[{\includegraphics[width=1in,height=1.25in,clip,keepaspectratio]{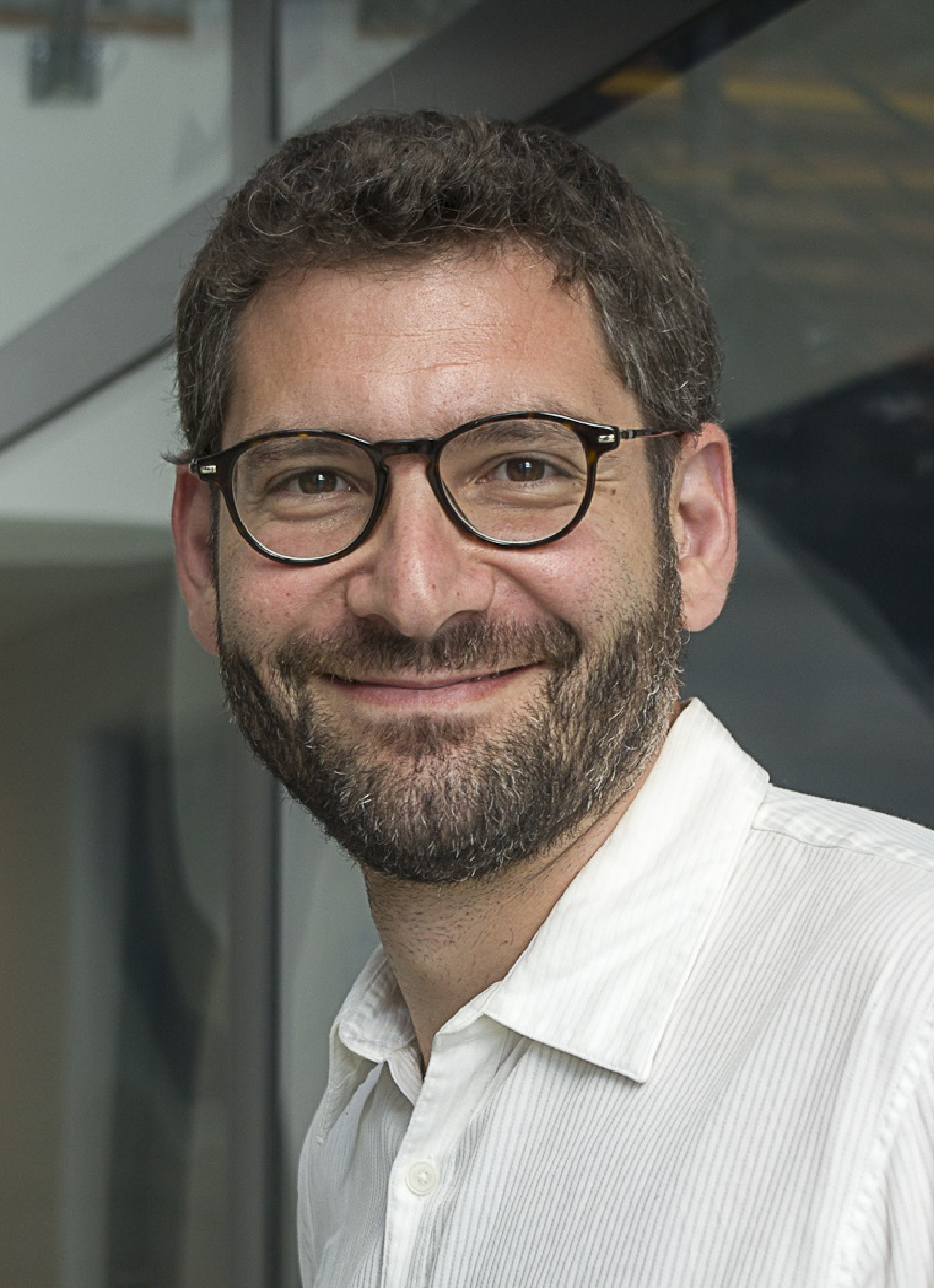}}]{Marios Kountouris (S’04–M’08–SM’15–F’23)} received the diploma degree in electrical and computer engineering from the National Technical University of Athens (NTUA), Greece in 2002 and the M.S. and Ph.D. degrees in electrical engineering from Télécom Paris, France in 2004 and 2008, respectively. He is currently a Professor at the Communication Systems department, EURECOM, Sophia-Antipolis, France. Prior to his current appointment, he has held positions at CentraleSupélec, France, the University of Texas at Austin, USA, Huawei Paris Research Center, France, and Yonsei University, South Korea. He is the recipient of a Consolidator Grant of the European Research Council (ERC) in 2020 on goal-oriented semantic communication. He has served as Editor for the IEEE Transactions on Wireless Communications, the IEEE Transactions on Signal Processing, and the IEEE Wireless Communication Letters. He has received several awards and distinctions, including the 2022 Blondel Medal, the 2020 IEEE ComSoc Young Author Best Paper Award, the 2016 IEEE ComSoc CTTC Early Achievement Award, the 2013 IEEE ComSoc Outstanding Young Researcher Award for the EMEA Region, the 2012 IEEE SPS Signal Processing Magazine Award, the IEEE SPAWC 2013 Best Paper Award and the IEEE Globecom 2009 Communication Theory Best Paper Award. He is an IEEE Fellow, an AAIA Fellow, and a chartered Professional Engineer of the Technical Chamber of Greece.
\end{IEEEbiography}

\begin{IEEEbiography}[{\includegraphics[width=1in,height=1.25in,clip,keepaspectratio]{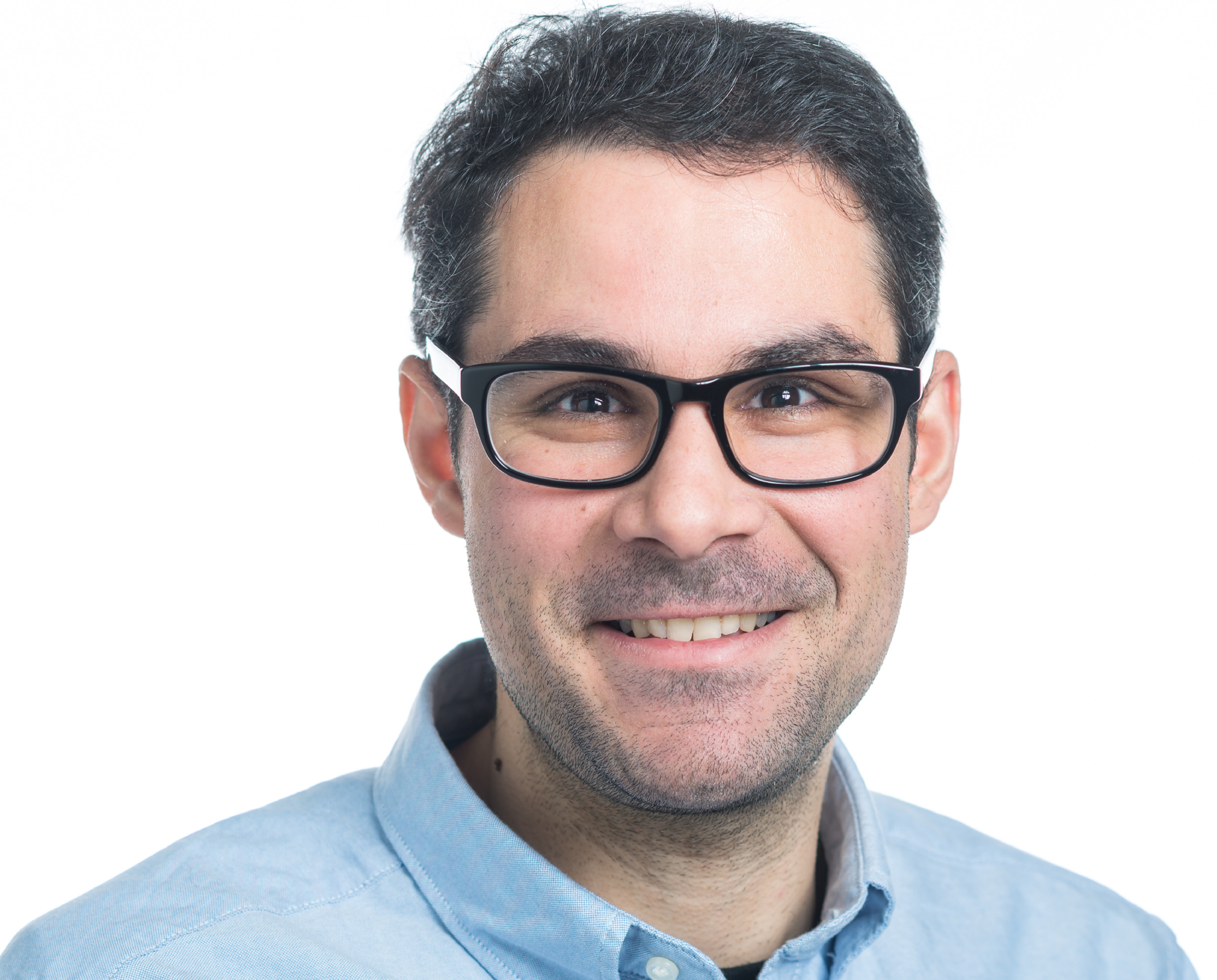}}]{Nikolaos Pappas}
	(Senior Member, IEEE) received a B.Sc. degree in computer science, a B.Sc. degree in mathematics, an M.Sc. degree in computer science, and a Ph.D. degree in computer science from the University of Crete, Greece, in 2005, 2012, 2007, and 2012, respectively. From 2005 to 2012, he was a Graduate Research Assistant with the Telecommunications and Networks Laboratory, Institute of Computer Science, Foundation for Research and Technology—Hellas, Heraklion, Greece, and a Visiting Scholar with the Institute of Systems Research, University of Maryland at College Park, College Park, MD, USA. From 2012 to 2014, he was a postdoctoral Researcher with the Department of Telecommunications, CentraleSupélec, Gif-sur-Yvette, France. He is currently an Associate Professor at the Department of Computer and Information Science at Linköping University, Linköping, Sweden. His main research interests include the field of wireless communication networks with an emphasis on semantics-aware communications, energy harvesting networks, network-level cooperation, age of information, and stochastic geometry. Dr. Pappas has served as the Symposium Co-Chair of the IEEE International Conference on Communications in 2022 and the IEEE Wireless Communications and Networking Conference in 2022. He is area editor of the IEEE OPEN JOURNAL OF THE COMMUNICATIONS SOCIETY and an Expert Editor for invited papers of the IEEE COMMUNICATIONS LETTERS. He is an Editor of the IEEE TRANSACTIONS ON MACHINE LEARNING IN COMMUNICATIONS AND NETWORKING and the IEEE/KICS JOURNAL OF COMMUNICATIONS AND NETWORKS. He is a guest editor of the IEEE NETWORK on "Tactile Internet for a cyber-physical continuum", and the IEEE IoT MAGAZINE on "Task-Oriented Communications and Networking for the Internet of Things". He has served as an Editor of the IEEE COMMUNICATIONS LETTERS and the IEEE TRANSACTIONS ON COMMUNICATIONS. He was a Guest Editor of the IEEE INTERNET OF THINGS JOURNAL on "Age of Information and Data Semantics for Sensing, Communication and Control Co-Design in IoT". 
\end{IEEEbiography}
\end{document}